\documentclass[journal]{IEEEtran}
\usepackage{url}
\usepackage{cite}
\usepackage{latexsym}
\usepackage{amsmath}
\usepackage{amstext,amsfonts,epsf,epsfig,pifont}
\usepackage{graphics,graphicx,bm,amsmath}

\newlength{\intwidth}

\newcommand{\sinc}{\ensuremath{\mathrm{sinc}}}

\newcommand{\W}{\ensuremath{W}}

\begin{document}

\noindent \emph{The following statements are placed here in accordance with the copyright policy of the Institute of Electrical and Electronics Engineers, Inc., available online at}
\url{http://www.ieee.org/web/publications/rights/policies.html}.\\

\noindent
Lilly, J. M., \&  Olhede, S. C. (2009). Higher-order properties of \indent analytic wavelets. \emph{IEEE Transactions on Signal Processing}, \indent \textbf{57} (1), 146--160.\\

\noindent This is a preprint version.  The definitive version is available from the IEEE or from the first author's web site, \url{http://www.jmlilly.net}.\\

\noindent \copyright 2009 IEEE. Personal use of this material is permitted. However, permission to reprint/republish this material for advertising or promotional purposes or for creating new collective works for resale or redistribution to servers or lists, or to reuse any copyrighted component of this work in other works must be obtained from the IEEE.\\

\newpage

\title{Higher-Order Properties of Analytic Wavelets}
\author{Jonathan~M.~Lilly,~\IEEEmembership{Member,~IEEE,}
and Sofia~C.~Olhede,~\IEEEmembership{Member,~IEEE}
\thanks{Manuscript submitted \today.  The work of J. M. Lilly was supported by a fellowship from the Conseil Scientifique of the Universit$\acute{\mathrm{e}}$ Pierre et Marie Curie in Paris, and by award \#0526297 from the Physical Oceanography program of the United States National Science Foundation. A collaboration visit by S.~C.~Olhede to Earth and Space Research in the summer of 2006 was funded by the Imperial College Trust.}
\thanks{J.~M.~Lilly is with Earth and Space Research, 2101 Fourth Ave., Suite 1310, Seattle, WA 98121, USA.}
\thanks{S.~C.~Olhede is with the Department of Statistical Science, University College London, Gower Street,
London WC1 E6BT, UK.}}

\markboth{IEEE Transactions on Signal Processing, Submitted February 2008}{Lilly \& Olhede: Higher-Order Properties of Analytic Wavelets}

\maketitle

\begin{abstract}
The influence of higher-order wavelet properties on the analytic wavelet transform behavior is investigated, and wavelet functions offering advantageous performance are identified. This is accomplished through detailed investigation of the generalized Morse wavelets, a two-parameter family of exactly analytic continuous wavelets.  The degree of time/frequency localization, the existence of a mapping between scale and frequency, and the bias involved in estimating properties of modulated oscillatory signals, are proposed as important considerations.  Wavelet behavior is found to be strongly impacted by the degree of asymmetry of the wavelet in both the frequency and the time domain, as quantified by the third central moments.  A particular subset of the generalized Morse wavelets, recognized as deriving from an inhomogeneous Airy function, emerge as having particularly desirable properties.  These ``Airy wavelets'' substantially outperform the only approximately analytic Morlet wavelets for high time localization.  Special cases of the generalized Morse wavelets are examined, revealing a broad range of behaviors which can be matched to the characteristics of a signal.
\end{abstract}
\begin{keywords}Wavelet transform, instantaneous frequency, ridge analysis, Hilbert transform, time-frequency analysis\end{keywords}
\IEEEpeerreviewmaketitle

\section{Introduction}

\IEEEPARstart{W}{avelet} analysis is a powerful and popular tool for the analysis of nonstationary signals.  The wavelet transform is a joint function of a time series of interest $x(t)$ and an analyzing function or wavelet $\psi(t)$. This transform isolates signal variability both in time $t$, and also in ``scale'' $s$, by rescaling and shifting the analyzing wavelet. The wavelet itself can be said to play the role of a lens through which a signal is observed, and therefore it is important to understand how the wavelet transform depends upon the wavelet properties.  This permits the identification of wavelets whose higher-order properties---with, say, duration held fixed---lead to the most accurate representation of the signal.

Here we focus on analytic, also known as progressive, wavelets---complex-valued wavelets with vanishing support on the negative frequency axis---defined in continuous time. Such wavelets are ideal for the analysis of modulated oscillatory signals, since the continuous analytic wavelet transform provides an estimate of the instantaneous amplitude and instantaneous phase of the signal in the vicinity of each time/scale location $(t,s)$. The analytic wavelet transform is the basis for the ``wavelet ridge'' method \cite{delprat92-itit,mallat}, which recovers time-varying estimates of instantaneous amplitude, phase, and frequency of a modulated oscillatory signal from the time/scale plane. On the other hand, the analytic wavelet transform can also be useful for application to very time-localized structures \cite{tu05-itit}, particularly if these features may appear as either locally even (symmetric) or locally odd (asymmetric) \cite{lilly03-pio}. The many useful features of analytic wavelets are covered in more depth by \cite{selesnick05-ispm}.

A promising class of exactly analytic wavelets is the generalized Morse wavelet family  \cite{daubechies88-ip}, the joint time/frequency localization properties of which were examined by \cite{olhede02-itsp}.  The generalized Morse wavelets have been used to estimate characteristics of a number of different non-stationary signals, including
blood-flow data \cite{olhede03c-itbe}, seismic and solar magnetic field data \cite{olhede03b-prsla}, neurophysiological time series \cite{brittain07-itbe}, and free-drifting oceanographic float records \cite{lilly06-npg}, and have also been utilized in image analysis \cite{antoine99-acha}.  With two free parameters,
the generalized Morse wavelets can take on a broad range of forms which has not yet been fully explored, and in fact this family encompasses most other popular analytic wavelets. Two parameters yield a natural characterization of an analytic wavelet function, since then the decay at both higher and lower frequencies from the center of the pass-band can be independently specified.  The generalized Morse wavelets will therefore be the focus of this study.

This work has three goals. The first goal, addressed in Section~2, is to identify important ways in which higher-order properties of analytic wavelets express themselves in the wavelet transform.  Third-order measures of the degree of asymmetry of the wavelet, both in the frequency domain and in the time domain, emerge as key quantities relating to the precise behavior of the wavelet transform;  we note that comparable behavior would be expected for discrete analytic wavelets at long time scales.  The second goal is then to establish the particular properties of the generalized Morse wavelets, and is accomplished in  Section~3.  A primary result is the existence of a sub-family, characterized by vanishing third derivative at the peak of the frequency-domain wavelet, which offers attractive behavior for the analysis of oscillatory signals.

The third goal, the focus of Section~4, is to provide practical guidelines for the choice of a continuous analytic wavelet appropriate for a particular task.  When calculating the wavelet transform, it is often desirable to choose wavelets to match the signal or structure of interest. The generalized Morse wavelets emerge as exhibiting a variety of possible behaviors, including limits in which the wavelet transform collapses to either the analytic filter or, in a certain sense, to the Fourier transform.  The sub-family mentioned above is shown to derive from an inhomogeneous Airy function, and we expect these ``Airy  wavelets'' will find value as superior alternatives to the popular Morlet wavelet.

All software related to this paper is distributed as a part of a freely available package of Matlab functions, called JLAB, available at the first author's website,  \url{http://www.jmlilly.net}.

\section{Analytic Wavelet Properties}
In this section, several desirable properties of continuous analytic wavelets are introduced: maximization of a conventional measure of the time/frequency energy concentration;  the existence of a unique relationship between scale and frequency; and minimization of the bias involved in estimating properties of oscillatory signals. The second and third of these will be shown to relate to the degree of asymmetry of the wavelet in the frequency and time domains, respectively, and more specifically to third order central moments.

\subsection{Definitions}
The continuous wavelet transform of a signal $x(t)\in L^2({\mathbb{R}})$ is a sequence of projections onto rescaled and translated versions of an analyzing function or ``wavelet'' $\psi(t)$,
\begin{eqnarray}
W(t,s)&\equiv& \int_{-\infty}^{\infty} \frac{1}{s} \psi^*\left( \frac{u-t}{s}\right) x(u)\,du\\& = &\frac{1}{2\pi}\int_{-\infty}^{\infty}\Psi^*(s\omega) X(\omega)\,e^{i\omega t}\,d\omega
\label{wavetrans}
\end{eqnarray}
where $\Psi(\omega)= \int_{-\infty}^\infty  \psi(t)e ^ {-i\omega t} \,dt$ is the Fourier transform of the wavelet and $X(\omega)$ is the Fourier transform of the signal. Note the choice of $1/s$ normalization rather than the more common $1/\sqrt{s}$, as we find the former to be more convenient for analysis of oscillatory signals. The wavelet is a zero-mean function which is assumed to satisfy the admissibility condition \cite{holschneider}
\begin{eqnarray}
c_{\psi}\equiv
\int_{-\infty}^\infty \frac{\left|\Psi(\omega)\right|^ 2}{\left|\omega\right|}\, d\omega&< & \infty\label{cpsidefinition}
\end{eqnarray}
and is said to be analytic if $\Psi(\omega)= 0 $ for  $\omega<0$. The wavelet modulus $\left|\Psi(\omega)\right|$ obtains a maximum at the \emph{peak frequency}  $\omega_\psi$; note that $\omega$ and $\omega_\psi$ are both radian frequencies.  It will also be convenient to adopt the convention that $\left|\Psi(\omega_\psi)\right|= 2$.

\subsection{An Overview of Analytic Wavelets}\label{analyticity}

A commonly used complex-valued wavelet is the Morlet wavelet \cite{holschneider}, which is essentially a Gaussian envelope modulated by a complex-valued carrier wave at radian frequency~$\nu$:
\begin{eqnarray}
\psi_{\nu}(t)&=& a_\nu \,  e^{-\frac{1}{2}\,t^2}\left[e^ {i\nu t}-e^ {-\frac{1}{2}\nu^2 }\right]\label{Morletwavelet}\\
\Psi_{\nu}(\omega)&=& a_\nu \,  e^{-\frac{1}{2}\,(\omega-\nu)^2}\left[1-e^ {-\omega\nu}\right].\label{Morletwaveletfrequency}
\end{eqnarray}
As the carrier wave frequency $\nu$ increases, more oscillations fit into the Gaussian window, and the wavelet becomes increasingly frequency-localized.  The second term in (\ref{Morletwavelet}) and (\ref{Morletwaveletfrequency}) is a correction necessary to enforce zero mean, while $a_\nu$ normalizes the wavelet amplitude. An expression for $a_\nu$, along with further details of the Morlet wavelet, is given in Appendix~\ref{Morletappendix}. The Morlet wavelet is not, however, exactly analytic---it is only approximately analytic for sufficiently large $\nu$, and this has important implications, as will be shown shortly.

A promising class of analytic wavelets are the generalized Morse wavelets, which have the frequency-domain form \cite{olhede02-itsp}
\begin{eqnarray}
\Psi_{\beta,\gamma}(\omega)&=& U(\omega) \,a_{\beta,\gamma}\, \omega^\beta e^{-\omega^\gamma}
\label{morse}
\end{eqnarray}
where $U(\omega)$ is the Heaviside step function and where
\begin{equation}
a_{\beta,\gamma}\equiv  2 (e\gamma/\beta)^{\beta/\gamma}\label{adef}
\end{equation}
is a normalizing constant. The peak frequency of these wavelets is given by $\omega_{\beta,\gamma}=
\left(\beta/\gamma\right)^{1/\gamma}$. The wavelet  $\psi_{\beta,\gamma}(t)$ is in fact the lowest-order member of an orthogonal family of wavelets for each ($\beta,\gamma$) pair \cite{olhede02-itsp},  but we will not be concerned with the higher-order members in the present paper. The generalized Morse wavelets are the solutions to a joint time/frequency localization problem, with analytic expressions for both the shape of the concentration region and the fractional energy concentration \cite{olhede02-itsp}. Note that we replace the subscript ``$\psi$'', denoting a property of an arbitrary analytic wavelet, with the subscript ``$\beta,\gamma$'' for specific properties of the generalized Morse wavelets.

The generalized Morse wavelets form a two-parameter family of wavelets, exhibiting an additional degree of freedom in comparison with the Morlet wavelet.  The nature of this additional degree of freedom has not yet been fully explored, but it would appear to control variation in higher-order wavelet properties with the time and frequency resolution held fixed.   Because of this adjustability, the generalized Morse wavelets can exhibit a very broad variety of behaviors, making (\ref{morse}) a fairly general prescription for constructing an exactly analytic wavelet. Furthermore, the generalized Morse wavelets subsume two other classes of commonly-used analytic wavelets, the analytic derivative of Gaussian wavelets \cite{tu05-itit} with $\gamma= 2$, and the Cauchy, also known as Klauder, wavelets \cite{holschneider} with $\gamma= 1$.

The number of other existing analytic continuous wavelets is fairly limited. There is also the complex Shannon wavelet [\citen{teolis},~p.~63]
\begin{eqnarray}
\psi_{S}(t)&=& \pi\, \sinc(t) \,  e^{i 2\pi t }\label{Shannonwavelet}\\
\Psi_{S}(\omega)&=&  2\,\mathrm{rect}  \left(\frac{\omega}{2\pi} +1 \right)\label{Shannonwaveletfrequency}
\end{eqnarray}
where $\sinc(t)$ is the sinc function and  $\mathrm{rect}(\omega)$ is the unit rectangle function.  The usefulness of the Shannon wavelet $\psi_{S}(t)$ is limited by its slow ($1/t$)  time decay, a consequence of the ``sharp edges'' of its Fourier transform. Another analytic wavelet is the Bessel wavelet \cite{lobos05-ee}, defined by
\begin{eqnarray}
\psi_{B}(t)&=& \frac{2}{\pi\sqrt{1 -it }}  \, K_1\left(2\sqrt{1 -it }\right) \label{Besselwavelet}\\
\Psi_{B}(\omega)&=&  2e^{-(\omega +1/\omega)}\label{Besselwaveletfrequency}
\end{eqnarray}
where  $K_1(\cdot)$ is the first modified Bessel function of the second kind. However these two wavelets are less commonly used than the Cauchy and Gaussian wavelets that are subsets of the generalized Morse family, and in addition, are restricted in their behaviors on account of having zero free parameters. These considerations justify our focus on the generalized Morse wavelets.

\subsection{Importance of Analyticity}\label{analyticitysection}

\begin{figure}[t]
\begin{center}
\hspace{-0.5cm}\includegraphics[width=3.5in,angle=0]{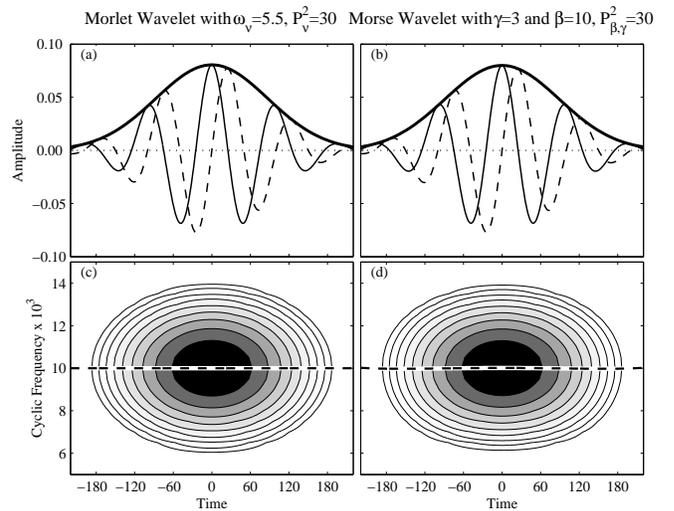}
\end{center}
            \caption{A Morlet wavelet (a), a generalized Morse wavelet (b), and their respective Wigner-Ville distributions (c,d). These two wavelets are fairly long in time, and in a sense that will be made precise later, these two wavelets can be said to have the same length. In (a) and (b), the thick solid, thin solid, and dashed lines correspond to the magnitude, real part, and imaginary part of the time-domain wavelet respectively. In panels (c) and (d), ten logarithmically spaced contours are drawn from the maximum value of the distribution to 1\% of that value. The thick dashed lines in (c) and (d) are the wavelet instantaneous frequencies.
            }\label{morsie_morlet_wigdist_long}
\end{figure}

The advantage of using precisely, as opposed to approximately, analytic wavelets such as the generalized Morse wavelets was demonstrated by \cite{olhede03a-prsla}, who showed that even small amounts of leakage to negative frequencies can result in spurious variation of the transform phase. It is important to emphasize this point for practical signal analysis.  As an example, a generalized Morse wavelet and a Morlet waveletcome together with their Wigner-Ville distributions \cite{mallat}, are shown in Figure~\ref{morsie_morlet_wigdist_long}.  The wavelet Wigner-Ville distribution is a fundamental time-frequency object which expresses the smoothing implicit in the wavelet transform; see \cite{mallat} for details. The instantaneous frequency \cite{boashash92a-ieee}---a quantity which reflects the time-varying frequency content of a modulated oscillatory signal---of each wavelet is also shown; in this case both instantaneous frequencies are very nearly constant, as would be the case for a sinusoid. Parameter settings have been chosen such that the width of the central time window (as measured by the standard deviation of the time-domain wavelet demodulated by its peak frequency, defined subsequently) in proportion to the period  $2\pi/\omega_\psi$ is the same for both wavelets. These two wavelets appear indistinguishable, and their Wigner-Ville distributions are nearly identical.

\begin{figure}[h]
\begin{center}
\hspace{-0.5cm}\includegraphics[width=3.5in,angle=0]{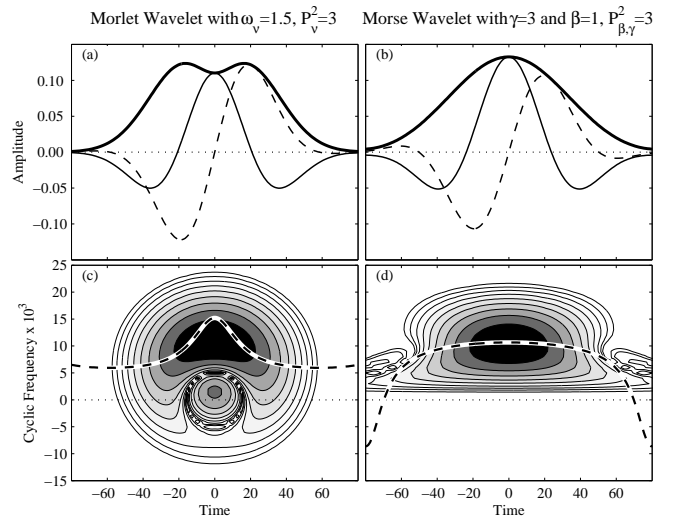}
\end{center}
            \caption{As with Figure~\ref{morsie_morlet_wigdist_long}, but for parameter settings giving wavelets that are very short in time.
            }\label{morsie_morlet_wigdist}
\end{figure}

However, if we narrow the time window of both wavelets by a factor of $\sqrt{10}$, in order to increase time resolution at the expense of frequency resolution, we obtain the wavelets shown in Figure~\ref{morsie_morlet_wigdist}. The Morlet wavelet now exhibits leakage to negative frequencies, as well as substantial instantaneous frequency fluctuations over the central window.  At this very narrow parameter setting, a local minimum has developed in the amplitude at the wavelet center on account of the correction terms.  By contrast, the generalized Morse wavelet has an instantaneous frequency which is nearly constant over the central window, and its Wigner-Ville distribution remains entirely concentrated at positive frequencies. Unlike the Morlet wavelet, the Morse wavelets remain analytic even for highly time-localized parameter settings. This is important for the analysis of strongly modulated functions, where the wavelets are required to be narrow in time to match the modulation timescale.

\begin{figure}[h!]
\begin{center}
\hspace{-0.5cm}\includegraphics[width=3in,angle=0]{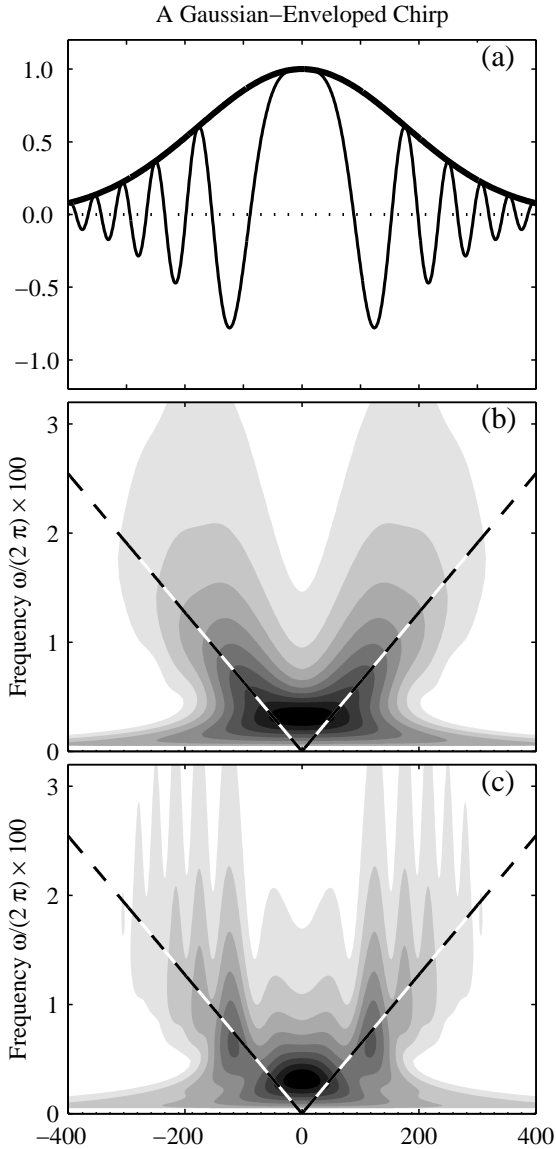}
\end{center}
            \caption{A Gaussian-enveloped chirp is shown in panel (a), while panels (b) and (c) show the wavelet transform of the signal with the generalized Morse wavelet and Morlet wavelet shown in Figure~\ref{morsie_morlet_wigdist}(a) and (b) respectively.  In (b) and (c), the derivative of the phase of the chirp signal is shown for reference as the dashed line.
            }\label{morsies_linear_chirp}
\end{figure}

The implications of the negative-frequency leakage of the Morlet wavelet's Wigner-Ville  distribution---a manifestation of its departure from analyticity---are drastically degraded transform properties. Figure~\ref{morsies_linear_chirp} shows the wavelet transform of a Gaussian-enveloped chirp using the two wavelets from Figure~\ref{morsie_morlet_wigdist}.  The chirp signal has a frequency which increases at a constant rate, passing through zero frequency at time $t=0$. The negative-frequency leakage of the Morlet wavelet leads to interference in the wavelet transform which accounts for its irregular structure.  Essentially this interference pattern can be understood as the interaction of the chirp with its image at negative frequencies. Estimates of amplitude or phase properties of the signal using the Morlet transform would be badly biased. Since the Morse wavelet has no support at negative frequencies, the interference is completely suppressed.

As an aside, we point out that in this example, the derivative of the phase of the chirp is shown for reference. However, this is not the same as the instantaneous frequency of the total signal. The maxima-line of the Morse wavelet transform follows the latter rather than the former; see  \cite{lilly08-itit} for further details.

This illustrates the importance of analyticity for the deterministic properties of the wavelet transform, the details of which have been investigated by other authors.  Statistical properties are also deteriorated by departures from analyticity: whereas an analytic transform applied to a Gaussian process produces Gaussian proper transform coefficients [\citen{olhede03a-prsla}, p.~424], this behavior is lost with a non-analytic analysis function.  Since the advantages of analytic wavelets are well established, our concern henceforth will be on the influence of analytic wavelet structure on transform properties.  We focus on Morse wavelets because they are a two parameter family encompassing most other major analytic wavelets. On account of its non-analyticity, the Morlet wavelet is not a valid point comparison in most of what follows.  We turn now to defining several important properties of analytic wavelets.

\subsection{Mapping Scale to Frequency}\label{frequencyinterpretation}
It is common practice to consider the scale $s$ as proportional to an inverse frequency. But it is critical to keep in mind that any assignment of frequency to scale is an interpretation, and there is in fact more than one valid interpretation. The ideal wavelet should have these different interpretations be identical, such that there is no ambiguity in assigning frequency to a given scale.

One may define three meaningful frequencies associated with the wavelet itself: the peak frequency $\omega_{\psi}$ at which the wavelet magnitude $|\Psi(\omega)|$ is maximized, which is also the mode of $|\Psi(\omega)|^2$; the  \emph{energy frequency}
\begin{eqnarray}
\widetilde \omega_{\psi}
& \equiv   & \frac{\int_0^\infty \omega |\Psi(\omega)|^2
  \,d\omega}{\int_0^\infty\left| \Psi(\omega)\right|^ 2
  \,d\omega}\label{energyfrequency}
\end{eqnarray}
which is the mean of $|\Psi(\omega)|^2$; and finally, the time-varying instantaneous frequency \cite{boashash92a-ieee} of the wavelet
\begin{eqnarray}
  \breve\omega_\psi(t)& = & \frac{d}{dt}\Im\left\{\ln\psi (t)\right\}=\frac{d}{dt}\arg\left\{\psi (t)\right\}\label{waveletinstantaneousfrequency}
\end{eqnarray}
evaluated at the wavelet center, $\breve\omega_\psi(0)$. A difference between $\omega_{\psi}$ and $\widetilde \omega_{\psi}$ is obviously an expression of frequency-domain asymmetry of the wavelet. On the other hand, the energy frequency and instantaneous frequency are related by \cite{cohen96-itsp}
\begin{equation}
\widetilde \omega_{\psi}=\frac{\int_{-\infty}^\infty\left|\psi (t)\right|^ 2\breve\omega_\psi(t)\,dt}{\int_{-\infty}^\infty\left|\psi (t)\right|^ 2 dt}
\end{equation}
and therefore a departure of $\breve\omega_\psi(0)$ from $\widetilde \omega_{\psi}$ implies that the wavelet frequency content is not uniform in time.

There correspond three separate interpretations of scale as frequency. Consider $x_o(t)=\cos\left(\omega_o t\right)$, having an analytic wavelet transform
\begin{equation}
W_o\left(t,s\right)
 = \frac{1}{2}\,\Psi^*\left(s\omega_o\right)e^{i\omega_o t}
\end{equation}
where the contribution from negative frequencies vanishes on account of the analyticity of the wavelet.  The scale at which the magnitude of the wavelet transform is maximum, obtained by solving
\begin{equation}
\frac{\partial}{\partial s}\left|W_o\left(t,s\right)\right|^ 2= 0,\label{amplituderidge}
\end{equation}
is found to be $s=s_\psi\equiv\omega_\psi/\omega_o$.  This is the same as the scale at which the rate of change of transform phase is equal to the signal frequency, that is, the scale at which
\begin{equation}
\frac{\partial}{\partial t}\Im\left\{\ln\left[W_o\left(t,s\right)\right])\right\}= \omega_o\label{phaseridge}
\end{equation}
is satisfied. The peak frequency $\omega_\psi$ therefore controls location of the amplitude maximum, and the rate of phase progression, of an oscillatory feature much broader in time than the wavelet.  Note that for more general signals (\ref{amplituderidge}) and (\ref{phaseridge}) respectively define the amplitude and phase ridge curves of the wavelet transform \cite{lilly08-itit}, from which the instantaneous frequency of the signal can be derived.

Secondly, one may form the energy-mean scale, which becomes for the sinusoid
\begin{equation}
\widetilde s_\psi \equiv  \frac{\int_{0}^{\infty}s \left|W_o\left(t, s\right)\right|^ 2 \,ds}{\int_{0}^{\infty} \left|W_o\left(t,s\right)\right|^ 2 \,ds}=\frac{\int_{0}^{\infty}s  \left|\Psi\left(s\omega_o\right)\right|^ 2 \,ds}{\int_{0}^{\infty}
\left|\Psi\left(s\omega_o\right)\right|^ 2\,ds}
\end{equation}
but following a change of variables, this is seen to be simply $\widetilde s_\psi= \widetilde\omega_\psi/\omega_o$. Thus $\widetilde\omega_\psi$ determines the scale at which the first moment of the modulus-squared wavelet transform of a sinusoidal signal occurs; this gives a integral measure of energy content of a signal across all scales.

Finally, let $W_\delta\left(t,s\right)$ be the wavelet transform of a Dirac delta-function $\delta(t)$ located at the origin. The rate of change of phase of this transform is
\begin{equation}
\frac{\partial}{\partial t}\Im\left\{\ln\left[W_\delta\left(t,s\right)\right])\right\}= \frac{1}{s}\,\breve\omega_\psi\left(\frac{t}{s}\right)\label{waveletinstantaneousfrequency}
\end{equation}
which, at the location of the delta-function, becomes  $\breve\omega_\psi\left(0\right)/s$. The wavelet central instantaneous frequency $\breve\omega_\psi\left(0\right)$ therefore controls the rate of phase propagation at the center of a feature much narrower than the wavelet.

We thus have that  $\omega_s\equiv  \omega_\psi/s$, $\widetilde \omega_s\equiv  \widetilde \omega_\psi/s$, and $\breve\omega_s \equiv \breve\omega_\psi\left(0\right)/s$  define three different mappings of scale to frequency. The first will correctly give the frequency of a pure sinusoid from the scale $s$ at which its transform obtains a maximum, the second will correctly give the frequency of a pure sinusoid from the energy-mean scale of the transform, and the third fixes the frequency to be the same as the phase progression of the transform at the location of an infinitesimally narrow impulse. As all of these are mappings arguably correct in different senses or for different types of signals, it is desirable that all three should be the same.  In that case there would be a unique and unambiguous interpretation of scale as frequency. It will be found that the $\gamma = 3$ generalized Morse wavelets have the frequency measures very nearly being equal, while maintaining exact analyticity as well as good time localization.

\subsection{Energy Localization}

The degree of energy localization of a wavelet is conventionally expressed in terms of its \emph{Heisenberg area} ~\cite{mallat}
\begin{eqnarray}
A_\psi& \equiv & \sigma_{t;\psi}\,\sigma_{\omega;\psi}\label{Heisenbergbox}
\end{eqnarray}
where the standard deviations
\begin{eqnarray}
\sigma_{t;\psi}^ 2& \equiv  &\omega_\psi^ 2 \,\frac{\int t^ 2|\psi(t)|^ 2\,dt}{\int|\psi(t)|^ 2\,dt}\label{timespread}\\
\sigma_{\omega;\psi}^ 2& \equiv  & \frac{1}{\omega_\psi^ 2} \, \frac{\int \left(\omega-\widetilde \omega_{\psi}\right)^2|\Psi(\omega)|^ 2\,d\omega}{\int | \Psi(\omega)|^ 2\,d\omega}\label{frequencyspread}
\end{eqnarray}
describe the wavelet spread in the time domain and frequency domain, respectively; here the standard deviations have been defined such that they are nondimensional. The Heisenberg area  $A_\psi$ obtains a minimum value of one-half for a function which has a Gaussian envelope, but such a function is not a wavelet because it is not zero-mean.

Note that there are other notions of time-frequency localization. The whole set of generalized Morse wavelets are optimally localized in that they maximize the eigenvalues of a joint time-frequency localization operator, as shown by \cite{daubechies88-ip} and investigated in further detail by \cite{olhede02-itsp}, and indeed this is the way the generalized Morse wavelets were initially constructed. However, the Heisenberg area is a valuable measure of time-frequency localization, since it is in standard usage and permits ready comparison among different functions. It will be shown later that the $\gamma = 3$ generalized Morse wavelets are close to the theoretical minimum of the Heisenberg area, while remaining exactly analytic, even for narrow time-domain settings; their concentration is comparable to or greater than that of the Morlet wavelet.

\subsection{Minimally Biased Signal Inference}\label{oscillatoryinference}

An important application of analytic wavelet analysis is to detect the properties of modulated oscillatory signals of the form \cite{delprat92-itit,mallat}
\begin{eqnarray}
x(t)&=& a_+(t) \cos \left[ \phi_+(t)\right].
\label {modulatedmodel}
\end{eqnarray}
The amplitude $a_+(t) > 0$ and phase  $\phi_+(t)$ in this model are uniquely defined in terms of the \emph{analytic signal} \cite{picinbono97-itsp}
\begin{eqnarray}
x_+(t)&\equiv &
2\int_{-\infty}^\infty U(\omega) X(\omega) \,e ^{i\omega t}\,d\omega
\label {analytictransform}
\end{eqnarray}
where $U(\omega)$ is again the Heaviside step function. In terms of the analytic signal, the original real-valued signal is written as
\begin{eqnarray}
x(t)&= & \Re\left\{x_+(t)\right\}= \Re\left\{a_+(t) e^ {i\phi_+(t)}\right\}
\label {analyticmodel}
\end{eqnarray}
and $a_+(t)$ and $\phi_+(t)$ are called the canonical amplitude and phase \cite{picinbono97-itsp}. The rates of change of the phase $\omega(t)\equiv d/dt\left\{\phi_+(t)\right\}$ and log-amplitude $\upsilon(t)\equiv d/dt\left\{\ln\left[a_+(t)\right]\right\}$ are called the instantaneous frequency and instantaneous bandwidth, respectively.

Wavelet ridge analysis \cite{delprat92-itit,mallat} is a second analysis step performed on an analytic wavelet transform which estimates the properties of the analytic signal or signals associated with the time series. The bias properties of wavelet ridge analysis were examined by \cite{lilly08-itit}, the results of which we make use of in this section. A dimensionless version of the wavelet frequency-domain derivative is defined by
\begin{eqnarray}
\widetilde\Psi_{n}(\omega)&\equiv& \omega^n\frac{\Psi^{(n)}(\omega)}{\Psi(\omega)}\label{normalizedwavelet}
\end{eqnarray}
where the superscript ``$(n)$'' denotes the $n$th-order derivative of $\Psi(\omega)$. With this definition, an exact form of the wavelet transform of an oscillatory signal, derived in \cite{lilly08-itit}, is
\begin{multline}
 x_{\psi}  (t) \equiv  \W\left(t, \omega_\psi/\omega(t)\right) \\=x_+(t)\times \left\{1 -\frac{1}{2}\left[  \frac{ a_+''(t)}{a_+(t)}+i\omega'(t) \right] \frac{\widetilde\Psi^*_{2}(\omega_\psi)}{\omega^ 2(t)}\right.\\\left.
  +\frac{i}{6} \left[\frac{a_+'''(t)}{a_+(t)}
  +3i\frac{a_+'(t)}{a_+(t)}\, \omega'(t) +i\omega''(t)\right]
\frac{\widetilde\Psi^*_{3}(\omega_\psi)}{\omega^ 3(t)}\right.\\\left.\label{locallyanalytic}
+\epsilon_{\psi,4}\left(t\right)\right\}
\end{multline}
where the quantity $\epsilon_{\psi,4}\left(t\right)$ is a bounded residual associated with truncating the integration outside of a finite range \cite{lilly08-itit}.

Equation (\ref{locallyanalytic}) defines a nonlinear smoothing of the analytic signal by the wavelet, resulting in a quantity $ x_{\psi}(t ) $ which departs from the analytic signal $ x_+  (t)$ on account of interactions between time-domain derivatives of the signal and frequency-domain derivatives of the wavelet. Note that these bias terms depend on wavelet derivatives evaluated only at the peak frequency $\omega_\psi$. Under a certain smoothness assumption on the original signal $ x (t)$, expected to hold when the signal is locally described as an oscillation, the term in (\ref{locallyanalytic}) associated with $\widetilde\Psi_{3}(\omega_\psi)$ is much smaller than that associated with $\widetilde\Psi_{2}(\omega_\psi)$, and the term associated with $\widetilde\Psi_{4}(\omega_\psi)$---here incorporated into the residual---is smaller still; see \cite{lilly08-itit} for details. For this reason it is more important to minimize $|\widetilde\Psi_{3}(\omega_\psi) |$ rather than $|\widetilde\Psi_{4}(\omega_\psi) |$ for a given $\widetilde\Psi_{2}(\omega_\psi)$.

Since, as shown below, the square root of normalized second derivative
$\left|\widetilde\Psi_{2}(\omega_\psi)\right|$ is a nondimensional measure of the wavelet duration, it appears desirable choose a wavelet with vanishing $\widetilde\Psi_{3}(\omega_\psi)$ for a fixed value of $\widetilde\Psi_{2}(\omega_\psi)$, as this removes the next-highest-order bias term for a given wavelet duration. It will be shown that the generalized Morse wavelets achieve this with the choice $\gamma = 3$. We are aware of no other analytic wavelets with this property, apart from the Shannon wavelets which we do not prefer on account of their poor time localization. The Bessel wavelets mentioned above have a very large value of the third derivative at the peak frequency, $\widetilde\Psi_{3}(\omega_\psi)= - 6$, which would result in substantial bias according to the analysis of \cite{lilly08-itit}.

\section{Properties of Generalized Morse Wavelets}
Having defined three useful properties of analytic wavelets---high time/frequency concentration, a unique interpretation of scale as frequency, and minimized bias for analyzing oscillatory signals---our next goal is to form explicit expressions of the relevant quantities for the generalized Morse wavelets.

\subsection{Frequency-Domain Moments}\label{momentsection}

The structure of a wavelet can be described in terms of its moments or cumulants.  We will use both the frequency-domain wavelet moments and the \emph{energy moments}
\begin{eqnarray}
M_{n;\psi}
& \equiv   & \frac{1}{2\pi}\int_0^\infty
  \omega^{n}\, \Psi(\omega) \,d\omega\\
N_{n;\psi}
& \equiv   & \frac{1}{2\pi}\int_0^\infty
  \omega^{n} \left|\Psi(\omega)\right|^ 2 \,d\omega
\end{eqnarray}
as well as the frequency-domain wavelet cumulants.  The wavelet moments are the terms in the Taylor series expansion
\begin{eqnarray}
\psi(t)& = &\sum_{n= 0} ^\infty\frac{(it)^n}{n!}\,M_{n;\psi}\label{psimoments}
\end{eqnarray}
while the coefficients $K_{n;\psi}$ in the expansion
\begin{eqnarray}
\ln\psi(t)& = &\sum_{n= 0} ^\infty\frac{(it)^n}{n!}\,K_{n;\psi}\label{psicumulants}
\end{eqnarray}
define the frequency-domain wavelet cumulants.  Technically, these are called ``formal moments'' and ``formal cumulants'' since the frequency-domain wavelet is not normalized as a probability density function.  The wavelet cumulants $K_{n;\psi}$ may be found in terms of the wavelet moments $M_{n;\psi}$ through $M_{0;\psi}=\exp\left(K_{0;\psi}\right)$ together with a recursion relation given in Appendix~\ref{momentappendix}.

The moments of the generalized Morse wavelets are
\begin{eqnarray}
M_{n;\beta,\gamma}
& = &\frac{a_{\beta,\gamma} }{2\pi\gamma}\Gamma\left(\frac{\beta+1+n}{\gamma}\right)\label{Morsemoment}
\end{eqnarray}
while the expression
\begin{eqnarray}
N_{n;\beta,\gamma}
& =   & \frac{2}{2 ^{(1+n)/\gamma}}\,M_{n;2\beta,\gamma}\label{energymoment}
\end{eqnarray}
gives the energy moments of the $(\beta, \gamma)$ wavelet in terms of the moments of the $(2\beta, \gamma)$ wavelet. Combining (\ref{Morsemoment}) with (\ref{recursion}) of Appendix~\ref{momentappendix}, one obtains simple expressions for the cumulants, the first three of which are given explicitly in Appendix~\ref{momentappendix}.  In Appendix~\ref{convergenceappendix} it is shown that for the generalized Morse wavelets, the series in (\ref{psimoments}) converges for
all $t$~such that $|t| < 1$ for $\gamma= 1$, while for $\gamma> 1$ the radius of convergence is infinite.

\subsection{Frequency-Domain Derivatives  / Time-Domain Moments}\label{timesection}
We will also need expressions for the dimensionless frequency-domain derivatives $\widetilde\Psi_n(\omega)$ evaluated at the peak frequency $\omega_{\psi}$, shown in Section~\ref{oscillatoryinference} to control the wavelet transform of an oscillatory signal.  The dimensionless derivatives can be cast in a form which is somewhat more straightforward to interpret. Let the time-domain moments of the wavelet demodulated by its peak frequency be denoted by
\begin{eqnarray}
m_{n;\psi}  &\equiv & \int_{-\infty}^\infty t^n\,e^{-i\omega_\psi t}\psi(t) \,dt
\label{Mdef}
\end{eqnarray}
and recall the correspondence between time-domain moments and frequency-domain derivatives [\citen{papoulis}, Section~5.5],
\begin{equation}
\frac{m_{n;\psi}}{m_{0;\psi}} =   i^n \frac{\Psi^{(n)}(\omega_\psi)}{\Psi(\omega_\psi)}=  i^n \frac{\widetilde\Psi_{n}(\omega_\psi)}{\omega_\psi^n}.
\end{equation}
Since $m_{1;\psi}= 0$, the $m_{n;\psi}$ are central moments of the demodulated wavelet $e^{-i\omega_\psi t}\psi(t)/m_{0;\psi}$. This suggests normalizing the higher-order demodulate moments by the second moment, as one would for a probability density function:
\begin{eqnarray}
\alpha_{n;\psi}&\equiv & \frac{\displaystyle\frac{m_{n;\psi}}{m_{0;\psi}}}{ \left[\displaystyle\frac{m_{2;\psi}}{m_{0;\psi}}\right]^{n/2}}
=i^n
\frac{\widetilde\Psi_{n}(\omega_{\psi})}{\left|\widetilde\Psi_{2}(\omega_{\psi})\right|^{n/2}}.
\end{eqnarray}
The normalized third and fourth central moments $\alpha_{3;\psi}$ and  $\alpha_{4;\psi}$ will be called the  \emph{demodulate skewness}  and  \emph{demodulate kurtosis}, owing to their formal resemblance to skewness and kurtosis of a probability density function. However, it is important to keep in mind that the demodulated wavelet is not a probability density function since it is in general complex-valued and not nonnegative.

Now define a dimensionless measure of the wavelet time-domain length or duration
\begin{equation}
P_\psi  \equiv  \pi \,  \frac{ 2\sqrt{ \frac{m_{2;\psi}}{m_{0;\psi}}}}{2\pi/\omega_{\psi}}
 = \omega_\psi \sqrt{ \frac{m_{2;\psi}}{m_{0;\psi}}}
 =\sqrt{\left|\widetilde\Psi_2(\omega_\psi)\right|}
\end{equation}
such that $P_\psi/\pi$ is the number of oscillations at the peak frequency which fit within the central wavelet window, as measured by the standard deviation of the demodulated wavelet. Increasing $P_\psi$ increases the frequency-domain curvature in the vicinity of the peak frequency, narrowing the wavelet in the frequency domain and hence broadening the wavelet in the time domain.

For the generalized Morse wavelets, we find in Appendix~\ref{derivativeappendix} that $P_{\beta,\gamma}$ is simply $\sqrt{\beta\gamma}$, while the demodulate skewness and kurtosis become
\begin{eqnarray}
\alpha_{3;\beta,\gamma}&=&i\frac{\gamma-3}{\sqrt{\beta\gamma}}=i  \frac{\gamma-3}{P_{\beta,\gamma}}\label{demodulateskewness}\\
\alpha_{4;\beta,\gamma}&=&
3 - \Im\left\{\alpha_{3;\beta,\gamma}\right\}^ 2 -\frac{2}{P_{\beta\gamma}^ 2}
\end{eqnarray}
where we note the demodulate skewness is a purely imaginary quantity. Note that the choice $\gamma= 3$ causes the demodulate skewness [and hence $\widetilde\Psi_3(\omega_\psi)$] to vanish---while simultaneously maximizing the magnitude of the demodulate kurtosis $\alpha_{4;\beta,\gamma}$ for a fixed value of $P_{\beta,\gamma}$.  As mentioned in Section~\ref{oscillatoryinference}, we are less concerned with the kurtosis than with the skewness: for signals which are locally oscillatory, it is expected that nonzero skewness will contribute more substantially to a bias of (\ref{locallyanalytic}), the analytic signal estimated from the wavelet transform \cite{lilly08-itit}.

\begin{figure*}[t]
\begin{center}
\includegraphics[width=4.5in,angle=-90]{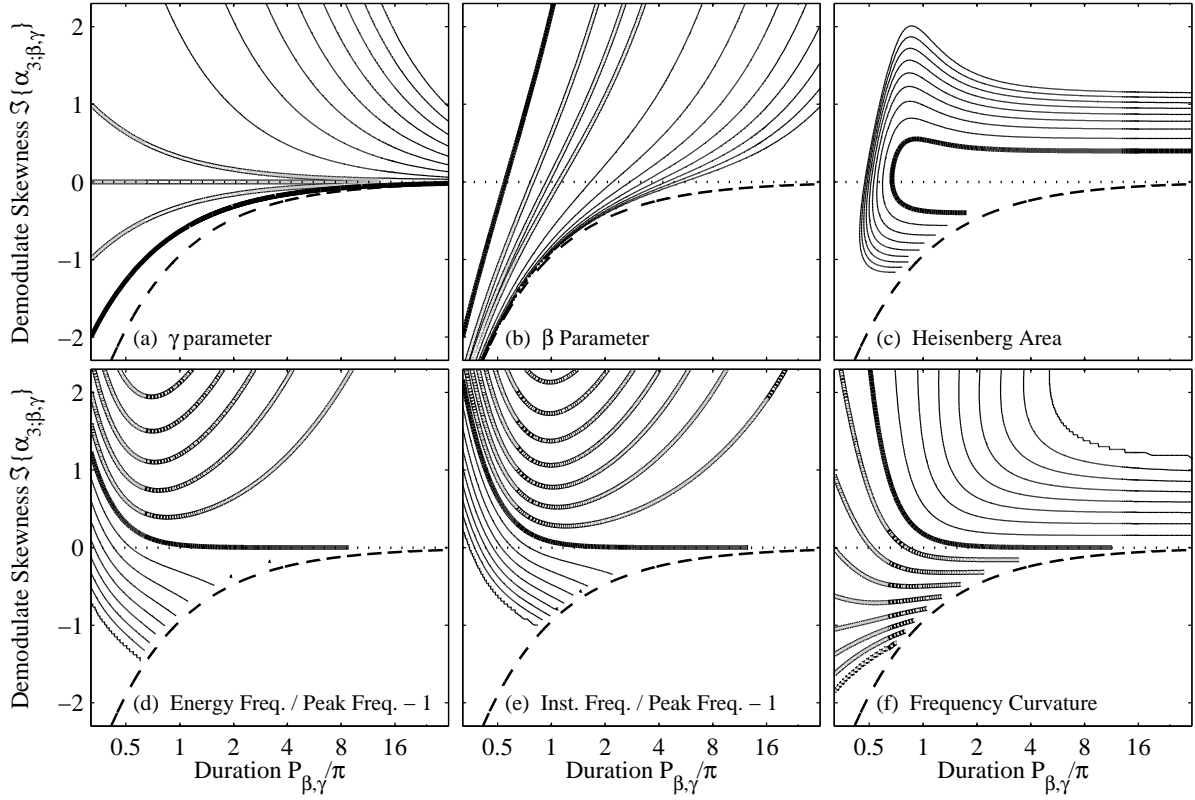}
\end{center}
            \caption{The behavior of Morse wavelet quantities as a function of $P_{\beta,\gamma}$ and the imaginary part of demodulate skewness $\alpha_{3;\beta,\gamma}$ is presented. All panels have the same axes, with    $P_{\beta,\gamma}/\pi$ being the x-axis and $\Re\left\{\alpha_{3;\beta,\gamma}\right\}$ being the y-axis. The six panels show six different quantities contoured as a function of this plane. Panels (a) and (b) show $\gamma$ and $\beta$, respectively.  Heavy solid lines show $\gamma= 1$ and $\beta= 1$, white lines with black outlines show $\gamma$ and $\beta= 2$, 3, and 4, and thin solid lines show $\gamma$ and $\beta= n ^2$ for integer $n$ with $3\le n\le 10$.  The dashed line in all panels is the $\gamma= 0$ contour. The Heisenberg area is shown in panel (c) with a contour interval of 0.01 from 0.51 to 0.59; the heavy solid line is the 0.51 contour. Panel (d) shows  $\widetilde \omega_{\beta,\gamma}/\omega_{\beta,\gamma}-1$, the difference of the ratio of the energy frequency to the peak frequency from unity, and similarly panel (e) shows $\breve\omega_{\beta,\gamma}(0)/\omega_{\beta,\gamma}-1$ where $\breve\omega_{\beta,\gamma}(0)$ is the value of the wavelet instantaneous frequency at the wavelet center. Finally panel (f) gives the dimensionless curvature of instantaneous frequency $\breve\omega_{\beta,\gamma}(t)$ as defined in (\ref{Morsecurve}).  The last three panels all have the same contours, which range from -0.2 to 0.2 with a contour integral of 0.025.  Thin solid contours are for positive values, white contours with black outlines are for negative values, and the heavy curve is for the zero contour. Note that $\gamma= 3$ lies along the x-axis.
            }\label{morsie_frequencies}
\end{figure*}

The dimensionless duration $P_{\beta,\gamma}$ and demodulate skewness $\Im\left\{\alpha_{3;\beta,\gamma}\right\}$ form a natural two-parameter description of the generalized Morse wavelets.  Particular values of $P_{\beta,\gamma}$ and $\Im\left\{\alpha_{3;\beta,\gamma}\right\}$ give a unique ($\beta,\gamma$) pair.  $P_{\beta,\gamma}$ is a normalized second-order moment, while $\alpha_{3;\beta,\gamma}$ is a normalized third-order moment measuring the degree of asymmetry of the demodulated wavelet in the time domain. For a given duration $P_{\beta,\gamma}$, a range of shapes can be obtained for the generalized Morse wavelets by adjusting $\Im\left\{\alpha_{3;\beta,\gamma}\right\}$. The fourth-order behavior, as expressed by the demodulate kurtosis $\alpha_{4;\beta,\gamma}$, is not free, but is an implicit function of the two lower-order quantities $P_{\beta,\gamma}$ and $\Im\left\{\alpha_{3;\beta,\gamma}\right\}$. This clarifies how $\beta$ and $\gamma$  translate directly into controlling the wavelet moments.

The generalized Morse wavelet parameters $\gamma$ and $\beta$ are plotted as a function of $P_{\beta,\gamma}$ and $\Im\left\{\alpha_{3;\beta,\gamma}\right\}$ in
Figure~\ref{morsie_frequencies}(a),(b). The parameter $\beta$ increases with increasing $P_{\beta,\gamma}$ but decreases with increasing $\Im\left\{\alpha_{3;\beta,\gamma}\right\}$. The $\gamma$ curves, on the other hand, change character at $\gamma= 3$, with contours of lower values being concave down and those of higher values being concave up; $\gamma= 3$ itself is the horizontal line $\Im\left\{\alpha_{3;\beta,\gamma}\right\}=0$. For a given $P_{\beta,\gamma}$ with $\beta\ge 0$ and $\gamma\ge 0$, $\Im\left\{\alpha_{3;\beta,\gamma}\right\}$ is bounded from below by $-3/P_{\beta,\gamma}$ for $\gamma= 0$, but has no upper bound; this lower bound is the cause of the empty region below the $\gamma= 0$ contour in these plots.

\subsection{Wavelet Frequency Measures}\label{frequencysection}
We now return to the question of assigning a frequency interpretation to the wavelets. From the results of the preceding sections, we see that the energy frequency defined in (\ref{energyfrequency}) becomes
\begin{eqnarray}
\widetilde \omega_{\psi}
& \equiv   & \frac{\int_0^\infty \omega \left|\Psi(\omega)\right|^2
  \,d\omega}{\int_0^\infty  \left|\Psi(\omega)\right|^2
  \,d\omega}= \frac{N_{1;\psi}}{N_{0;\psi}}\label{Omegatilde}
\end{eqnarray}
while the time-varying instantaneous wavelet frequency (\ref{waveletinstantaneousfrequency}) is [using (\ref{psicumulants})]
\begin{eqnarray}
  \breve\omega_\psi(t)& = & \frac{d}{dt}\Im\left\{\ln\psi (t)\right\}=K_{1;\psi} -\frac{1}{2}K_{3;\psi}t^ 2 + \ldots
\end{eqnarray}
when expanded in terms of the wavelet frequency-domain cumulants; only the odd cumulants appear because of taking the imaginary part. Thus $\breve\omega_\psi(0)=K_{1;\psi}=M_{1;\psi}/M_{0;\psi}$.  The quantity
\begin{equation}
\frac{1}{K_{2;\psi}^{3/2}}\frac{d^ 2\breve\omega_{\psi}}{dt^ 2}(0) \label{Morsecurve} =-\frac{K_{3;\psi}}{K_{2;\psi}^{3/2}}
\end{equation}
is a nondimensional measure of the curvature of the instantaneous frequency evaluated at the wavelet center. The instantaneous frequency curvature has a simple interpretation for a real-valued, nonnegative definite frequency domain wavelet such as a generalized Morse wavelet.  For such a wavelet, $\Psi(\omega)/M_{0;\psi}$ is a probability density function having $K_{3;\psi}/K_{2;\psi}^{3/2}$ as its coefficient of skewness, which is the negative of the dimensionless instantaneous frequency curvature. Thus frequency-domain skewness corresponds to curvature of the instantaneous frequency.

One finds for the generalized Morse wavelets that the energy frequency is
\begin{equation}
\widetilde \omega_{\beta,\gamma} =   \frac{N_{1;\beta,\gamma}}{N_{0;\beta,\gamma}}=\frac{1}{2 ^{1/\gamma}} \label{Morseenergyfrequency} \frac{M_{1;2\beta,\gamma}}{M_{0;2\beta,\gamma}}
 =   \frac{1}{2 ^{1/\gamma}} \frac{\Gamma\left(\frac{2\beta+2}{\gamma}\right)}{\Gamma\left(\frac{ 2\beta+1}{\gamma}\right)}
\end{equation}
while
\begin{equation}
\breve\omega_{\beta,\gamma}(0) =   K_{1;\beta,\gamma}= \frac{M_{1;\beta,\gamma}}{M_{0;\beta,\gamma}}=\label{Morsefreak} \frac{\Gamma\left(\frac{\beta+2}{\gamma}\right)}{\Gamma\left(\frac{\beta+1}{\gamma}\right)}
=2^{1/\gamma}\widetilde \omega_{\beta/2, \gamma}
\end{equation}
gives the time-varying instantaneous frequency at the wavelet center. An expression for the frequency curvature may be found from (\ref{Morsemoment}) together with (\ref{cumulant1}) and (\ref{cumulant3}) of Appendix~\ref{momentappendix}.

The behaviors of these frequency measures of the generalized Morse wavelets versus  $P_{\beta,\gamma}/\pi$  and $\Im\left\{\alpha_{3;\beta,\gamma}\right\}$ are shown in Figure~\ref{morsie_frequencies}(d),(e),(f). A change in character is observed at  $\gamma= 3$.  At $\gamma= 3$, the ratio of the energy frequency to the peak frequency $\widetilde \omega_{\beta, \gamma}/ \omega_{\beta, \gamma}$, and the ratio of the instantaneous frequency at the wavelet center $(t= 0)$ to the peak frequency $\breve \omega_{\beta, \gamma}(0)/ \omega_{\beta, \gamma}$, are both very close to unity, except for very short wavelet durations with $P_{\beta,\gamma}/\pi<1$.  For larger values of $\gamma$, or positive $\Im\left\{\alpha_{3;\beta,\gamma}\right\}$, both of these two ratios are generally smaller than unity, while for smaller values of $\gamma$ or negative $\Im\left\{\alpha_{3;\beta,\gamma}\right\}$ both ratios generally exceed unity. The exception is for very short durations  $P_{\beta,\gamma}/\pi<1$, where one finds these ratios becoming increasingly large and positive as $P_{\beta,\gamma}/\pi$ decreases with fixed $\Im\left\{\alpha_{3;\beta,\gamma}\right\}$.  Meanwhile the instantaneous frequency curvature at the wavelet center,
 Figure~\ref{morsie_frequencies}(f), exhibits a similar pattern but with the sign reversed.  The $\gamma= 3$ wavelets have negligible instantaneous frequency curvature except as the duration becomes very short.

For $P_{\beta,\gamma}/\pi>1$, the $\gamma= 3$ wavelets have $\widetilde \omega_{\beta, \gamma}\approx \omega_{\beta, \gamma}\approx \omega_{\beta, \gamma}(0)$ to a very good approximation, nearly obtaining the unambiguous interpretation of scale as frequency which was sought in Section~\ref{frequencyinterpretation} while remaining exactly analytic.
That $\omega_{\beta, \gamma}$ and $\widetilde \omega_{\beta, \gamma}$ should be almost identical for some value of $\gamma$ is not obvious.  The former is a simple algebraic expression in terms of powers of $\beta$ and $\gamma$, whereas the latter is given by a ratio of gamma functions. In Appendix~\ref{gammaappendix} it is shown that for increasing $\beta$, the ratio of these two quantities converges rapidly to unity for $\gamma=3$ due to the asymptotic behavior of the gamma function.

\begin{figure*}[tb]
\begin{center}
\hspace{-1in}\includegraphics[width=5in,angle=-90]{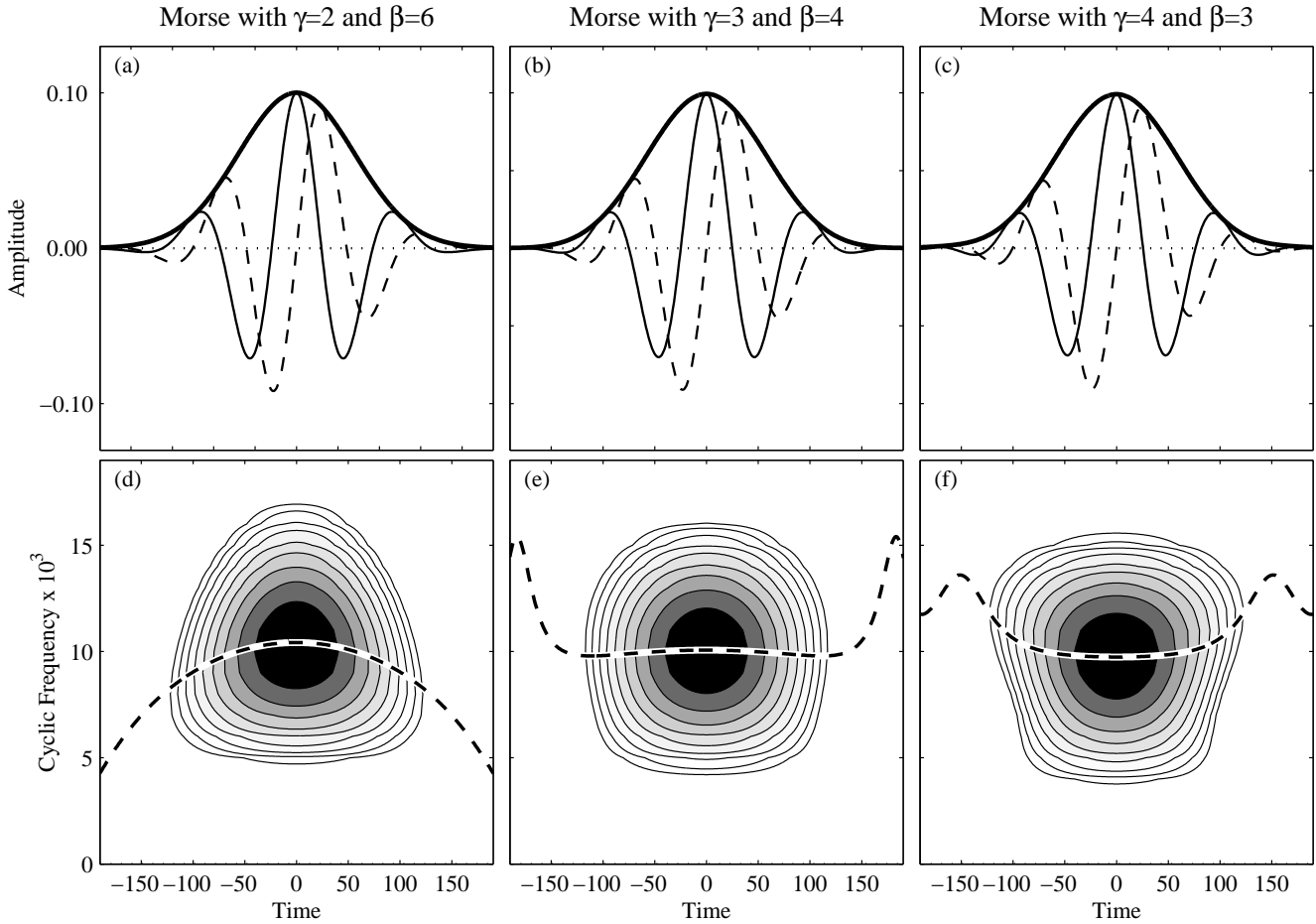}
\end{center}
            \caption{The time-domain generalized Morse wavelets (a,b,c) and their Wigner-Ville distributions (d,e,f).  All three wavelets have the same value of
             $P_{\beta,\gamma}(\omega_{\beta,\gamma}) =\sqrt{\beta\gamma} = 2\sqrt{3}$, but differing values of $\gamma$ as indicated in the captions.  Line styles and contour intervals are as in Figure~\ref{morsie_morlet_wigdist}.}
 \label{morsie_wigdist_three}
\end{figure*}

The sign change of the wavelet frequency curvature observed in Figure~\ref{morsie_frequencies}(f) gives the border between two qualitatively different behaviors, as is seen in Figure~\ref{morsie_wigdist_three}. For negative curvature, we have \emph{concave}  wavelets in which the instantaneous frequency takes on its maximum value at the wavelet center, while for positive curvature the wavelets are  \emph{convex} and the instantaneous frequency takes on a minimum value at the wavelet center.  Regions of large amplitude thus correspond to regions of high frequency for the concave case, but to regions of low frequency for the convex case.  The degree of convexity, or concavity, controls how the wavelet filter will respond preferentially to signals having frequency minima, or maxima, at the wavelet center.  For $P_{\beta,\gamma}/\pi>1$, the $\gamma= 3$ wavelets are the division between these two cases.  The $\gamma= 3$ wavelets have very small instantaneous frequency curvature, and have Wigner-Ville distributions which are roughly symmetric about the central frequency, as in Figure~\ref{morsie_wigdist_three}(b). It was seen earlier in
Figure~\ref{morsie_morlet_wigdist}(d) that this symmetry becomes compromised for very time-localized settings, corresponding to ``squashed'' appearance of the Wigner-Ville distribution and to the curvature apparent in Figure~\ref{morsie_frequencies}(f) for small $P_{\beta,\gamma}/\pi$.

\subsection{Energy Localization}

We now address the problem of time/frequency concentration as measured by the Heisenberg area. The frequency spread defined in (\ref{frequencyspread}) simplifies to
\begin{eqnarray}
\sigma_{\omega;\psi}^ 2\label{Morsefrequencyspread}
& = & \frac{1}{\omega_\psi^2} \left[\frac{N_{2;\psi}}{N_{0;\psi}}-\widetilde \omega_{\psi}^ 2\right]
\end{eqnarray}
making use of the definition of $\widetilde \omega_{\psi}$  (\ref{Omegatilde}). For the time-domain spread (\ref{timespread}), note that one may show
\begin{eqnarray}
\sigma_{t;\psi}^ 2
& = & \omega_\psi^2\frac{\int |\Psi'(\omega)|^ 2\,d\omega}{\int | \Psi(\omega)|^ 2\,d\omega}
\end{eqnarray}
using the relation between time-moments and frequency-domain derivatives  together with Parseval's theorem. This becomes for the generalized Morse wavelets
\begin{multline}
\frac{\sigma_{t;\beta,\gamma}^ 2}{\omega_{\beta, \gamma}^2}=\frac{a_{\beta,\gamma}^ 2}{N_{0; \beta, \gamma}}\times \\
\left[\beta^ 2\frac{N_{0; \beta- 1, \gamma}}{a_{\beta-1,\gamma}^ 2}+
\gamma^ 2\frac{N_{0; \beta- 1+ \gamma, \gamma}}{a_{\beta-1+ \gamma,\gamma}^ 2}
-2\beta\gamma\frac{N_{0; \beta- 1+ \gamma/2, \gamma}}{a_{\beta-1+ \gamma/2,\gamma}^ 2}
\right]
\end{multline}
which can then be expressed using (\ref{adef}) and (\ref{energymoment}).

The Heisenberg area is shown in Figure~\ref{morsie_frequencies}(c) as a function of the wavelet duration $P_{\beta,\gamma}/\pi$ and demodulate skewness
$\Im\left\{\alpha_{3;\beta,\gamma}\right\}$.  It is clear the wavelets exhibiting small time-domain skewness have a small Heisenberg area.  As $P_{\beta,\gamma}/\pi$ increases, the Heisenberg area approaches the limiting value of one-half for any value of the demodulate skewness $\Im\left\{\alpha_{3;\beta,\gamma}\right\}$.  The theoretical minimum value of the Heisenberg area of one-half is not quite obtained, except in the limit of long duration, evidently on account of asymmetry induced by the constraint of analyticity.  However, we may point out that perfect concentration is not achieved by the Morlet wavelet either---although constructed from a Gaussian, the existence of the correction terms lead to departure from the theoretical minimum value of one-half.  In fact, numerical computations we have performed (not presented here) show the Heisenberg area of the $\gamma= 3$ wavelet is comparable to or smaller than that of the Morlet wavelet.

Summarizing the results in this section, we see from Figure~\ref{morsie_frequencies} the special properties of the $\gamma= 3$ wavelets. In the vicinity of $\gamma= 3$, the generalized Morse wavelets obtain their minimum Heisenberg area, and for sufficiently large duration $P_{\beta,\gamma}$ the peak frequency $\omega_{\beta,\gamma}$, energy frequency $\widetilde \omega_{\beta,\gamma}$, and central instantaneous frequency $\breve \omega_{\beta,\gamma}(0)$ all become indistinguishable while the wavelet instantaneous frequency curvature (\ref{Morsecurve}) vanishes.  The $\gamma=3$ generalized Morse wavelets therefore obtain the ideal behavior with respect to the three issues raised in the Section~2 for $P_{\beta,\gamma}/\pi>1$. For extremely time-concentrated wavelets with $P_{\beta,\gamma}/\pi<1$, the curves along which the frequency pairs are identical, the frequency curvature vanishes, and the Heisenberg area is minimized, all begin to diverge from one another.

\section{Special Cases of Generalized Morse Wavelets}

In this section we step back from an emphasis on properties relevant for analysis of oscillatory signals.  Instead we examine the broad variety of behavior of the generalized Morse wavelets, with the idea in mind that these could be considered a generic family of analytic wavelets appropriate for analyzing many different types of signals.  In this section we therefore isolate special cases of these wavelets, exploring the boundaries of the family as well as the relationships among its different members.

\subsection{Interpretations of $\beta$ and $\gamma$}
We discuss two important interpretations of $\beta$ and $\gamma$, the first pertaining to the relationships among different members of the generalized Morse wavelet family, and the second to the time-domain and frequency-domain decay of the wavelets.
\subsubsection{Differentiation and Warping\label{warpingsection}}

The $(\beta,\gamma)$ generalized Morse wavelet was represented as a nonlinear transformation of the $\left(\frac{\beta+\frac{1}{2}}{\gamma}-\frac{1}{2},1\right)$ wavelet by \cite{olhede02-itsp}. This representation was crucial for deriving the localization properties of the generalized Morse wavelets in the time-frequency plane for $\beta>\frac{\gamma-1}{2}>0$, since the time-frequency localization operator for which these wavelets form the eigenvectors \cite{daubechies88-ip} is only well defined in this case.  Here we present an alternate construction which more directly reflects the different roles of the $\beta$ and $\gamma$ parameters.

Note that over the entire range $\gamma\geq 0$ and $\beta\geq 0$, the inverse Fourier transform $\psi_{\beta,\gamma}(t)$ of (\ref{morse}) still defines a valid filtering function. We refer to $\psi_{\beta,\gamma}(t)$ over this entire range as the \emph{generalized Morse filter}, only a subset of which corresponds to the generalized Morse wavelets. In fact, it is easy to see that this filter is zero-mean for $\beta> 0$. Computing  $c_{\psi}$ defined in (\ref{cpsidefinition}) for the generalized Morse wavelets, we find [using (\ref{Morsemoment})]
\begin{eqnarray}
c_{\beta,\gamma}
&= &\frac{a_{\beta,\gamma}^ 2}{\pi\gamma \,2 ^ {2\beta/\gamma+1}}\Gamma\left(\frac{2 \beta}{\gamma}\right)
\end{eqnarray}
and over the entire range $\gamma>0$ and $\beta>0$ admissibility is satisfied, as is the constraint of finite energy. Thus only $\beta= 0$ or $\gamma= 0$ are not valid wavelets.

Note that the time-domain $\beta= 0$, $\gamma>1$ filter
\begin{eqnarray}
\psi_{0,\gamma}(t)&=&\frac{1}{\pi} \int_{-\infty}^{\infty} e^{-\omega^\gamma} e^ {i\omega t}\,d\omega
\end{eqnarray}
may be expressed in terms of the frequency-domain $\beta= 0$, $\gamma=1$ filter as
\begin{eqnarray}
\psi_{0,\gamma}(t)&= &\frac{1}{2\pi} \int_{-\infty}^{\infty} \Psi_{0,1}(\omega^\gamma)\,e^ {i\omega t}\,d\omega\label{warping}
\end{eqnarray}
since $\Psi_{0,1}(\omega) =  2e^{-\omega}$. Equation (\ref{warping}) states that the frequency-domain power distribution of $\Psi_{0,1}(\omega)$ is mapped onto different Fourier components through the substitution $\omega\mapsto\omega^ \gamma$.  But substituting the inverse Fourier transform, $\Psi_{0,1}(\omega)
 =  \int_{-\infty}^{\infty}\psi_{0,1}(t)e^{-i\omega t}\,dt$, one may write instead
\begin{eqnarray}
\psi_{0,\gamma}(t)
& = & \int_{-\infty}^{\infty} \psi_{0,1}(u)K_{\gamma}(t,u) \,du
\end{eqnarray}
where we have defined
\begin{eqnarray}
K_{\gamma}(t,u)& \equiv  &  \frac{1}{2\pi} \int_{0}^{\infty}e^ {i\omega t-i\omega^\gamma u}\,d\omega
\end{eqnarray}
as a time-domain transformation kernel function. Note that for  $\gamma= 1$  one has $K_{1}(t,u)=\delta(t-u)$ where $\delta(t)$ is the Dirac delta-function. Thus incrementing $\gamma$ is accomplished by a frequency-domain warping.

Subsequently, the $\beta\geq 1$ filter is obtained from the $\beta= 0$ filter for fixed $\gamma$ and $\beta\in{\mathbb{N}}$ via the time-domain differentiation
\begin{eqnarray}
\psi_{\beta,\gamma}(t)\label{betaderivative}
& = & a_{\beta,\gamma}(-i) ^\beta\frac{1}{2} \frac{d^\beta}{d t^\beta}\psi_{0,\gamma}(t).
\end{eqnarray}
Therefore all generalized Morse wavelets can be generated by beginning with $\Re\{\psi_{0,1}(t)\}$, making this function analytic to obtain $\psi_{0,1}(t)$, warping the frequency content to increment~$\gamma$, and then differentiating in the time domain to increment~$\beta$.  The nature of the originating function $\Re\{\psi_{0,1}(t)\}$ will be seen shortly.

\subsubsection{Frequency and Time Decay}
Clearly the  parameter $\gamma$ also controls the high-frequency decay of the wavelet. We now show that $\beta$ controls the time-domain decay.  The time-domain form of the generalized Morse wavelets is expressed by the inverse Fourier transform
\begin{equation}
\psi_{\beta,\gamma}(t)=\frac{1}{2\pi}\int_0^{\infty}a_{\beta,\gamma} \, \omega^{\beta}e^{-\omega^{\gamma}}
e^{i\omega t}\;d\omega.\label{timedomain}
\end{equation}
One may obtain their asymptotic time domain behavior using the method of [\citen{wong80-siam},~p.~407] by noting that
\begin{equation}
\omega^{\beta}e^{-\omega^{\gamma}}=\sum_{s=0}^{\infty}
\frac{(-1)^{s}}{s!}\omega^{\gamma s+\beta}.
\end{equation}
Inserting this into (\ref{timedomain}), we find that the integrals of the terms in this summation, while possibly divergent, are \emph{Abel summable} [\citen{wong80-siam},~p.~407]
and it follows from this reference that
\begin{multline}
\psi_{\beta,\gamma}(t)=a_{\beta,\gamma}\sum_{s=0}^{\infty}
\frac{(-1)^{s}}{s!} \exp\left\{\frac{i\pi
(s\gamma+\beta+1)}{2}\right\}\times \\\frac{\Gamma(s\gamma+\beta+1)}{t^{s\gamma+\beta+1}}.
\end{multline}
We therefore obtain the asymptotic behavior
\begin{equation}
\psi_{\beta,\gamma}(t)\sim a_{\beta,\gamma} \,e^ {i\pi(\beta +1)/2}\frac{\Gamma(\beta+1)}{t^{\beta +1}},\,\,\,\,\,\,\,|t|\longrightarrow\infty
\end{equation}
since the smallest power of $1/t$ dominates at large times.   The $O(t^ {-(\beta+1)})$ behavior could have been anticipated from the fact that the frequency-domain wavelet $\Psi_{\beta,\gamma}(\omega)$ is $\beta$ times differentiable but has a singularity in the $(\beta+1)$st derivative at $\omega= 0$.

\begin{figure*}[tb]
\begin{center}
\includegraphics[width=4.5in,angle=-90]{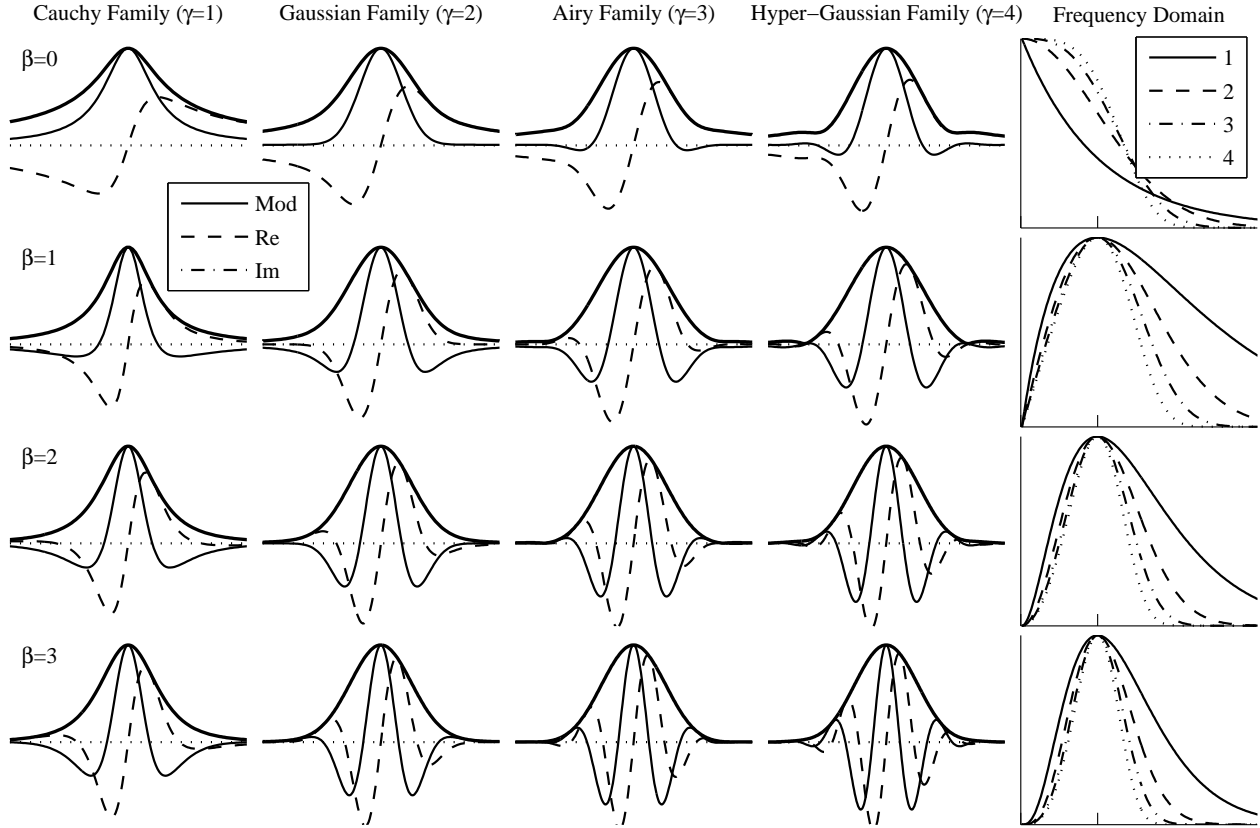}
\end{center}
            \caption{The time-domain forms of the generalized Morse filters for $\gamma= 1$--4 and $\beta=0$--3 are shown in the first four columns, while the fifth column shows the frequency-domain version of filters with $\gamma= 1$--4 for each value of $\beta$, with line styles as labeled. All wavelets with $\beta\neq0$ have the time axis normalized by $P_{\beta,\gamma}$, while the $\beta=0$ filters have been rescaled along the time axis such that their first time-domain derivatives (multiplied by a constant) are the $\beta=1$ wavelets shown.
            }\label{morsie_families}
\end{figure*}
The different roles of $\beta$ and $\gamma$ are illustrated in Figure~\ref{morsie_families}, which shows the first sixteen generalized Morse wavelet filters at integer $\beta\geq0$ and $\gamma\geq1$. Note that the appearance of the twelve wavelets ($\beta\neq0$) varies dramatically despite the fact that all have been set to have the same peak frequency. The action of differentiation (increasing $\beta$) is to broaden the central portion of the filter, while at the same time making the long-time decay more rapid.  On the other hand, increasing $\gamma$ reduces the curvature of the filter envelope at its center, also causing it to broaden, but without changing the long-time decay. This broadening of the central window width as $\beta$ or $\gamma$ increases agrees with our earlier identification of $P_{\beta\gamma}=\sqrt{\beta\gamma}$ as a dimensionless time-domain duration. Adjusting $\beta$ and $\gamma$ together therefore permits the ``inner'' width of the wavelet window to be controlled independently from the long-time decay.

\subsubsection {Symmetry Versus Compactness}
Earlier it was shown that time-domain symmetry of the demodulated wavelet is controlled by $\gamma$ through the ``demodulate skewness'' parameter $\alpha_{3;\beta,\gamma}=i (\gamma -3)/P_{\beta,\gamma}$ (\ref{demodulateskewness}).  Thus we can interpret $\beta$ as the decay, or compactness, parameter and $\gamma$ as the symmetry parameter.  Note the differing behaviors of the wavelet with fixed  $P_{\beta,\gamma}=\sqrt{\beta\gamma}$ on either side of $\gamma=3$. For $\gamma\leq 3$, time decay increases as $\beta$ increases from a minimum at $\beta=P_{\beta,\gamma}^ 2/3$, and the corresponding decrease in $\gamma$ makes the wavelet less symmetric.   On the other hand, for $\gamma\geq 3$, decreasing $\beta$ from a maximum at $\beta=P_{\beta,\gamma}^ 2/3$ also makes the demodulated wavelet less symmetric.  In this case the wavelet is most symmetric when its time decay is also strongest, and this occurs at $\gamma=3$.  Time-domain symmetry and compactness are therefore antagonistic for $\gamma< 3$ but covary for $\gamma> 3$.

\subsection{Domain Boundaries\label{Fourier}}
Next we examine the behaviors of the generalized Morse wavelets at extreme values on the $(\beta,\gamma)$ plane.

\subsubsection{The Analytic Filter Family\label{analytic}}
The frequency-domain generalized Morse filter for $\beta= \gamma= 0$ is simply twice the unit step function
\begin{eqnarray}
\Psi_{0,0}(\omega)&=& 2 U(\omega)
\end{eqnarray}
so that the time-domain Morse filter is the analytic filter \cite{poletti97-itsp}
\begin{eqnarray}
\psi_{0,0}(t)&=&
\delta(t)+
\frac{i}{\pi t}
\end{eqnarray}
which, of course, is not a wavelet. Application of this filter to a signal $x (t)$ simply recovers the analytic version of the signal, i.e.
\begin{eqnarray}
W_{x;0,0}(t,s)&= & x_+ (t),
\end{eqnarray}
independent of scale. This shows that the generalized Morse filter includes the analytic filter as a special case.  Commuting the differentiation operator with the analytic filter in (\ref{betaderivative}) shows that taking the wavelet transform with the wavelet $\psi_{\beta,0}(t)$ for $\beta\in{\mathbb{Z}}$ essentially involves taking the  $\beta$th derivative of the analytic signal.

\subsubsection{The Complex Exponential Limit\label{Fourier}}
The generalized Morse wavelets have an interesting behavior in the case $\beta\longrightarrow\infty$.  It follows from asymptotic expression for gamma function ratio (\ref{Morsegammaapproximation}) in Appendix~\ref{gammaappendix} that
\begin{eqnarray}
\frac{M_{n;\beta,\gamma}}{M_{0;\beta,\gamma}}&\sim&\left(\beta/\gamma\right)^{n/\gamma}
=\left[\omega_{\beta,\gamma}\right]^{n},\,\,\,\,\,\beta\longrightarrow\infty.\label{momentlimit}
\end{eqnarray}
Now, the moment expansion of the wavelet  (\ref{psimoments}) can be rewritten as
\begin{eqnarray}\label{demodulatedexpansion}
\psi_{\beta,\gamma}(t)/M_{0;\beta,\gamma}&=&\sum_{n=0}^\infty\frac{\left(it\right)^n}{n!}\frac{M_{n;\beta,\gamma}}{M_{0;\beta,\gamma}}
\end{eqnarray}
while a complex sinusoid at the wavelet peak frequency has a Taylor-series expansion
\begin{eqnarray}\label{sinusoidexpansion}
e^{i\omega_{\beta,\gamma}t}&=&\sum_{n=0}^\infty\frac{\left(it\right)^n}{n!}\left[\omega_{\beta,\gamma}\right]^{n}.
\end{eqnarray}
From (\ref{momentlimit}) we then see that for any fixed $n$, the $n$th moment of the normalized wavelet $\psi_{\beta,\gamma}(t)/M_{0;\beta,\gamma}$ becomes identical with the $n$th moment of the complex sinusoid $e^{i\omega_{\beta,\gamma}t}$ as $\beta$  approaches infinity. In this sense the generalized Morse wavelets approach a complex sinusoid as $\beta$ increases.  Equating (\ref{demodulatedexpansion}) and (\ref{sinusoidexpansion}) over some range of times would however require careful consideration of terms with $n=O(\beta)$  in the summations.

\subsection{The  $\gamma$ Families}
Finally we examine in more detail the generalized Morse wavelet families for the first few integer values of $\gamma$.

\subsubsection{The Cauchy Wavelets \label{cauchies}}
The $\gamma=1$ family
corresponds to the Cauchy wavelets [\citen{holschneider},~p.~28--29].  For $\beta>0$ and $\gamma=1$ the generalized Morse filter becomes the  \emph{analytic Cauchy filter}
\begin{eqnarray}
\psi_{0,1}(t)&=&\frac{1}{\pi}\int_0^{\infty} e^{-\omega}e^{i\omega t}\;d\omega=
\frac{1}{\pi(1-i t)}\\&=&\frac{1}{\pi(1+ t^2)}+i\frac{t}{\pi(1+ t^2)}
\end{eqnarray}
such that $\Re\left\{\psi_{0,1}(t)\right\}$ is the Witch of Agnesi curve, or, equivalently, the standard Cauchy probability distribution. This filter therefore specifies the joint effect of applying the analytic filter and smoothing by the Witch of Agnesi.  The $\beta$th filter for $\beta\in{\mathbb{N}}$ and $\beta\geq 1$ is then obtained by (\ref{betaderivative}) to be
\begin{eqnarray}
\psi_{\beta,1}(t)&=&\left(e/\beta\right)^\beta\frac{1}{\pi}\frac{\Gamma(\beta+1)}{(1-i t)^{\beta+1}},
\label{cauchy}
\end{eqnarray}
a form which, it turns out, is in fact valid for all $\beta>0$ \cite{olhede02-itsp}, not just the integers.

Following the results of Section~\ref{warpingsection}, all generalized Morse filters with $\beta\in{\mathbb{N}}$ and $\gamma\geq 1$ are generated from the Witch of Agnesi
(the real-valued curve in the upper of left-hand corner of Figure~\ref{morsie_families}) by analytization followed by warping followed by differentiation. Although not itself a wavelet, this time- and frequency-localized function forms the basis for all generalized Morse wavelets, and can therefore be thought of as the ``queen mother wavelet'' function.  The next two subsections demonstrate that the $\gamma=2$ and  $\gamma=3$ warpings generate two other important functions, the Gaussian probability density function and the inhomogeneous Airy function.

\subsubsection{The Analytic Derivative of Gaussian Wavelets
\label{analgaus}}
The $\gamma=2$ family
corresponds to analytic Derivative of Gaussian wavelets \cite{tu05-itit}. With $\beta=0$ and $\gamma = 2$, the generalized Morse filter becomes
\begin{eqnarray}
\psi_{0,2}(t)&=&\frac{1}{\pi}\int_0^{\infty}\label{Gaussianwavelet}
e^{-\omega^{2}}e^{i\omega t}\;d\omega\\&=&\frac{1}
{2\sqrt{\pi}}\left[e^{-t^2/4}+i\frac{2}
{\sqrt{\pi}}\,D(t/2)\right]
\end{eqnarray}
where $D\left(t\right)\equiv e^{-t^2}\int_{0}^{t} e^{u^2}\,du$ is the Dawson function. This extends the representation
of [\citen{olhede02-itsp},~p.~2667] which is only valid for the real part and for even values of $\beta$.  The $(\beta,2)$ wavelets are given for integer $\beta>0$, with $H_{\beta}(x)$ denoting the $\beta$th Hermite polynomial [\citen{abramowitz72},\;eqn.~22.2.14], by
\begin{multline}
\psi_{\beta,2}(t)=\frac{a_{\beta,2}}{4\sqrt{\pi}}\left(\frac{i}{2}\right)^{\beta}\times \\
\left\{H_{\beta}\left(t/2\right)e^{-t^2/4}+i(-1)^\beta \frac{2}
{ \sqrt{\pi}}\,D^ {(\beta)}(t/2)\right\}
\end{multline}
using (\ref{betaderivative}) and where
\begin{multline}
D^ {(n)}(t)= (-1)^n\times \\\left\{H_{n}\left(t
\right)D(t)-\sum_{k=1}^{n}\begin{pmatrix} n\\ k
\end{pmatrix}
H_{n-k}\left(t\right)i^{k-1}H_{k-1}\left(it
\right)\right\}\label{Dawsonderivatives}
\end{multline}
gives the form of the $n$th derivative of the Dawson function, which has been derived using Leibniz's theorem [\citen{abramowitz72},\;eqn.~3.3.8].

These wavelets have been proposed for singularity analysis by \cite{tu05-itit},
but no analytic expression for their time-domain form has been given previously as far as the authors are aware.  As illustrated in Figure~\ref{morsie_wigdist_three}(d), the instantaneous frequency curve for the analytic Derivative of Gaussian wavelets is concave; this is true for all  $\beta$ as Figure~\ref{morsie_frequencies}(c) shows. Their frequency domain behavior makes them less appropriate for the analysis of oscillations than the $\gamma= 3$ wavelets.

\subsubsection{The Airy Wavelets \label{airy}}
The $\gamma= 3$ generalized Morse wavelets in fact derive from an inhomogeneous Airy function, therefore we suggest calling this family the \emph{Airy wavelets}. The second inhomogeneous Airy function $\mathrm{Hi}(z)$, also known as the second Scorer function, is defined by the integral [\citen{abramowitz72},~p.~448,\;eqn.~10.4.44]
\begin{eqnarray}
\mathrm{Hi}(z)& \equiv  &  \frac{1}{\pi}\int_0^{\infty} e^{-u^3/3}e^{zu }\,du.
\end{eqnarray}
Thus the generalized Morse filter with $\beta= 0$ and $\gamma= 3$ is simply
\begin{equation}
\psi_{0,3}(t)=\frac{1}{\pi}\int_0^{\infty} e^{-\omega^{3}}e^{i\omega t}\;d\omega=
\frac{1}{3^ {1/3}}\,\mathrm{Hi}\left(\frac{it}{3^ {1/3}}\right)
\end{equation}
which is the inhomogeneous Airy function evaluated at an imaginary argument.  Differentiating the analytic Airy filter $\psi_{0,3}(t)$ $\beta$ times, as in (\ref{betaderivative}), one obtains
\begin{eqnarray}
\psi_{\beta,3}(t)&=&
a_{\beta,3}(-i)^{\beta}\frac{1} {2}\frac{1}{3^ {1/3}}\frac{d^{\beta}}{dt^{\beta}}\,\left[\mathrm{Hi}\left(\frac{it}{3^ {1/3}}\right)\right]
\end{eqnarray}
as an expression for the $\beta$th Airy wavelet $\psi_{\beta,3}(t)$ with $\beta\in{\mathbb{N}}$ and $\beta\geq 1$. Note that the $\beta= 1$ Airy wavelet is not within the localization regime $\beta>(\gamma -1)/2$. Examples are shown in Figure~\ref{morsie_morlet_wigdist}(b) and Figure~\ref{morsie_wigdist_three}(b).
As already discussed, the instantaneous frequency
of the wavelet is nearly constant over the width of the wavelet, and the
wavelet function exhibits no preference for its Wigner-Ville distribution to skew to smaller or larger frequencies on its periphery.
\subsubsection{The Hypergaussian Wavelets
\label{hypergaus}}
The $\gamma=4$ family does not have an analytic time-domain expression in terms of known functions as far as the authors are aware. However, this family is interesting because it is the first integer $\gamma$ family exhibiting convex behavior of the instantaneous frequency curve.  We may note that the analytic Gaussian filter is generated from the analytic Cauchy filter via the frequency-domain warping
\begin{eqnarray}
\psi_{0,2}(t)&= &\frac{1}{2\pi} \int_{-\infty}^{\infty} \Psi_{0,1}(\omega^2)\,e^ {i\omega t}\,d\omega
\end{eqnarray}
while the analytic $\gamma=4$ filter may be expressed as
 \begin{eqnarray}
\psi_{0,4}(t)&= &\frac{1}{2\pi} \int_{-\infty}^{\infty} \Psi_{0,2}(\omega^2)\,e^ {i\omega t}\,d\omega.
\end{eqnarray}
Thus the relation between $\psi_{0,4}(t)$ and $\psi_{0,2}(t)$ is the same as that between $\psi_{0,2}(t)$ and $\psi_{0,1}(t)$.  We therefore suggest ``Hypergaussian'' as a name for $\psi_{0,4}(t)$ since it involves a second iteration of the nonlinear operation creating the analytic Gaussian filter from the analytic Cauchy filter.

\section {Discussion and Conclusions}

This paper has examined the higher-order properties of analytic wavelets and their impact on the behavior of the wavelet transform.  Three important wavelet properties were discussed---time-frequency localization in terms of the Heisenberg area, the existence of a unique correspondence between scale and frequency, and minimized bias in the extraction of oscillatory signals. The latter two were shown to be related to third order moments of the wavelet. These properties were examined for the generalized Morse wavelets, a two-parameter family of exactly analytic wavelets.

The existence of a unique correspondence between scale and frequency
requires symmetry about the peak frequency, as measured by the frequency-domain skewness, and also equality between the mean and the mode of the squared modulus of the frequency-domain wavelet. Minimized bias in estimating instantaneous properties of modulated oscillatory signals was found to require a vanishing third derivative at the wavelet peak frequency, which is equivalent to a vanishing third central moment of the time-domain demodulated wavelet. Thus with a lower-order property held fixed---such as the wavelet duration in proportion to its period,  denoted here by $P_\psi$---choosing a wavelet which has a high degree of symmetry in both the time domain and the frequency domain leads to good performance for the analysis of oscillatory signals. These results for continuous analytic wavelets could also contribute to an improved understanding of the behavior of discrete analytic wavelets, such as those of \cite{selesnick05-ispm}.

One member of the generalized Morse wavelet family was found to have zero asymmetry in the time domain, as measured by the third central moment of a demodulated version of itself, as well as competitive performance in terms of other criteria.  This is the $\gamma= 3$ wavelet, shown herein to be derived from an inhomogeneous Airy function.  In fact the Airy wavelet preserves the spirit of the Morlet wavelet more than the Morlet itself, remaining nearly symmetric in the frequency domain and maintaining a nearly optimal Heisenberg area even at high time concentration, yet without compromising its exact analyticity.

The roles of the two parameters $\gamma$ and $\beta$ in setting practical properties of the wavelet filters was investigated in detail. Here we showed that the former controls the width of the inner wavelet window without impacting the time decay, while increasing the latter broadens the wavelet central window but increases the rate of decay at large times. The generalized Morse wavelets include as special cases the Cauchy wavelets ($\gamma= 1$) as well as analytic versions of the Derivative of Gaussian wavelets ($\gamma= 2$). The Airy wavelets emerge as the approximate boundary between two qualitatively different sorts of behavior, which we identify as convex or concave depending upon the sense of curvature of instantaneous frequency curve.  The broad range of behavior of the generalized Morse wavelets, together with their attractive properties for certain values of $\beta$ and $\gamma$, suggests their use as a generic family of exactly analytic wavelets.

\appendices

\section {The Morlet Wavelet}\label{Morletappendix}

In this section we address some details of the Morlet wavelet (\ref{Morletwavelet}--\ref{Morletwaveletfrequency}). First we find an expression for the peak frequency $\omega_\nu$ at which the frequency-domain wavelet obtains its maximum value, which is not the same as the carrier frequency $\nu$.  The peak frequency $\omega_\nu$ of the Morlet wavelet occurs where the first derivative
\begin{eqnarray}
\Psi_{\nu}'(\omega)&=& a_\nu \,  e^{-\frac{1}{2}\,(\omega-\nu)^2}\left[\nu+\omega\left(e^{-\omega\nu}-1 \right)\right]\label{Morletwaveletfrequency1}
\end{eqnarray}
vanishes.  This occurs when
\begin{equation}
\omega-\nu=\omega e^{-\omega\nu}\label{Morletvanishes}
\end{equation}
the solution to which may be found by introducing $\widetilde \nu \equiv \nu/\omega_\nu$.  Then (\ref{Morletvanishes}) leads to
\begin{equation}
\omega_\nu(\widetilde\nu) = \sqrt{-\frac{\ln\left(1-\widetilde\nu\right)}{\widetilde\nu}}\label{Morletequation}
\end{equation}
and since $0<\nu<\omega_\nu$, we can numerically solve (\ref{Morletequation}) on the interval $0<\widetilde \nu<1$ to obtain $\omega_\nu(\widetilde\nu)$. Setting
\begin{equation}
a_\nu \equiv 2\frac{\omega_\nu}{\nu}\,e^{\frac{1}{2}(\omega_\nu-\nu)^2} \label{Morletnormalization}
\end{equation}
for the normalization function $a_\nu$ obtains our chosen value of $\Psi_\nu(\omega_\nu)=2$, and from the above parametric form for $\omega_\nu$ we likewise know $a_\nu$ as a function of the carrier wave frequency.

Additional wavelet properties are given by the value of higher-order derivatives at the peak frequency.  One may verify
\begin{equation}
\Psi_{\nu}''(\omega)= -\left[\omega(\omega-\nu)+1\right]\Psi_{\nu}(\omega) -\left[2\omega-\nu\right]\Psi_{\nu}'(\omega) \label{Morletwaveletfrequency2}
\end{equation}
as an expression for the second derivative in the frequency domain, which leads to
\begin{eqnarray}
\widetilde\Psi_{\nu}^{(2)}(\omega_\nu)&=& -\omega_\nu^2\left[\omega_\nu(\omega_\nu-\nu)+1\right]
 \end{eqnarray}
for the normalized second derivative evaluated at the peak frequency. The wavelet duration is then
\begin{eqnarray}
P_\nu\equiv \sqrt{-\widetilde\Psi_{\nu}^{(2)}(\omega_\nu)}&=& \omega_v\sqrt{\omega_\nu(\omega_\nu-\nu)+1} \label{MorletwaveletP}
\end{eqnarray}
and one may note that as $\nu$ becomes large, one has $\omega_\nu\sim\nu$, $\widetilde\Psi_{\nu}^{(2)}(\omega_\nu)\sim-\nu^2$, and $P_\nu\sim \nu$.

\section {Wavelet Moments and Cumulants}\label{momentappendix}

To find the relation between the wavelet cumulants and the moments, note that
\begin{multline}
\psi(t) = \exp\left(\ln\left[\psi(t)\right]\right)=\exp\left(\sum_{n=0}^\infty \frac{(it)^n}{n!}K_{n;\psi}\right)\\=e^{K_{0;\psi}}
\left[1+\sum_{n= 1}^\infty \frac{(it)^n}{n!}B_n\left(K_{1;\psi},K_{2;\psi},\ldots K_{n;\psi}\right) \right]\label{expcumulants}
\end{multline}
[using (\ref{psicumulants})] which implicitly defines $B_n\left(c_1,c_2,\ldots  c_n\right)$, the $n$th-order complete Bell polynomial; see \cite{lilly08-itit} and references therein for details. Then equating powers of $t$ between (\ref{psimoments}) and (\ref{expcumulants}), one finds the moments are given in terms of the cumulants as $M_{0;\psi}=\exp\left(K_{0;\psi}\right)$ and
\begin{eqnarray}
\frac{M_{n;\psi}}{M_{0;\psi}}& = &B_n\left(K_{1;\psi},K_{2;\psi},\ldots K_{n;\psi}\right) \quad n\geq 1.\label{momentcumulant}
\end{eqnarray}
Inverting (\ref{momentcumulant}) leads to
\begin{eqnarray}
K_{1;\psi}& =&\frac{M_{1;\beta,\gamma}}{M_{0;\beta,\gamma}}\label{cumulant1}\\
K_{2;\psi} &=&\frac{M_{2;\psi}}{M_{0;\psi}}-\frac{M_{1;\psi}^ 2}{M_{0;\psi}^ 2}\label{cumulant2} \\
K_{3;\psi} &=& \frac{M_{3;\psi} }{M_{0;\psi}}-3 \label{cumulant3} \frac{M_{1;\psi}}{M_{0;\psi}}\frac{M_{2;\psi}}{M_{0;\psi}}+2\frac{M_{1;\psi} ^ 3}{M_{0;\psi}^ 3}
\end{eqnarray}
as expressions for the first three cumulants. These differ from the usual expressions between moments and cumulants [e.g. \citen{wiki:cumulant}] because the frequency-domain wavelet is not normalized as a probability density function, that is, $M_{0;\psi}\neq1$. More generally, the recursion relation
\begin{eqnarray}
K_{n;\psi}&  = & \frac{M_{n;\psi}}{M_{0;\psi}}  -\sum_{k = 1} ^ {n -1} \left(\!\!\begin{array}{c}n-1\\k -1\end{array}\!\!\right) K_{k;\psi} \frac{M_{n-k;\psi}}{M_{0;\psi}}\label{recursion}
\end{eqnarray}
relates the moments to the cumulants and vice-versa.

\section {Convergence of Morse Moment Expansion}\label{convergenceappendix}

Here we investigate the convergence of the moment expansion (\ref{psimoments}) for the generalized Morse wavelets.  The series converges for
all $t$ such that $|t| < r$, where $r$ is a positive constant referred to as the
radius of convergence [\citen{howie}, p. 203]. This radius may be determined by the
ratio test as
\begin{eqnarray}
r^{-1}& = &  \lim_{n\longrightarrow\infty} \frac{n!}{(n +1)!}\frac{M_{n+1;\psi}}{M_{n;\psi}}
\end{eqnarray}
and one finds for the generalized Morse wavelets with fixed $(\beta,\gamma)$ that
\begin{eqnarray}
r^{-1}& = &  \lim_{n\longrightarrow\infty} \frac{1}{n +1}\frac{\Gamma\left(\frac{\beta+1 +n+1}{\gamma}\right)}{\Gamma\left(\frac{\beta+1 +n}{\gamma}\right)}\\
&= &\lim_{n\longrightarrow\infty} n^{1/\gamma -1}\left(1/\gamma\right)^ {1/\gamma}\end{eqnarray}
using (\ref{Morsemoment}) and the asymptotic behavior of the gamma function given subsequently in (\ref{Morsegammaapproximation}).  The moment expansion for the generalized Morse wavelets therefore has radius of convergence $r=1$ for $\gamma= 1$, and infinite radius of convergence for $\gamma> 1$.

\section{Wavelet Frequency-Domain Derivatives}\label{derivativeappendix}
To find the generalized Morse wavelet frequency-domain derivatives, first note that there exists a simple expression
\begin{equation}
\omega ^ {n}\frac{d^n}{d\omega^n} \ln\Psi_{\beta,\gamma}(\omega)= \left[(-1)^{n-1}(n-1)!\right]\beta
-\omega^\gamma\prod_{p=0}^{n-1}\left(\gamma-p\right)
\end{equation}
[$n\ge 1$] for the derivative of the logarithm of the wavelet.
Taylor-expanding the frequency-domain generalized Morse wavelet about any fixed frequency $\omega_o$ leads to
\begin{equation}
\frac{\Psi_{\beta,\gamma}(\omega)}{\Psi_{\beta,\gamma}(\omega_o)}=
1+\sum_{n=1}^\infty \frac{(\omega/\omega_o- 1)^n}{n!} \,\widetilde \Psi_{n;\beta,\gamma}(\omega_o)
\end{equation}
but at the same time
\begin{multline}
\Psi_{\beta,\gamma}(\omega)=e^ {\ln\Psi_{\beta,\gamma}(\omega)} =\Psi_{\beta,\gamma}(\omega_o)\times \\\exp\left (
1+\sum_{n= 1}^\infty \frac{(\omega/\omega_o- 1)^n}{n!} \,\omega_o^n\frac{d ^n}{d\omega ^n}\left.\ln\Psi_{\beta,\gamma}(\omega) \right|_{\omega=\omega_o}\right)
\end{multline}
and therefore,  using (\ref{expcumulants})  and equating terms we obtain
\begin{multline}
\widetilde\Psi_{n;\beta,\gamma}(\omega)=B_n\left(\omega \frac{d}{d\omega} \label{Morsewaveletderivatives} \ln\Psi_{\beta,\gamma}(\omega), \omega^2\frac{d^2}{d\omega^2} \ln\Psi_{\beta,\gamma}(\omega),\right.\\\left.\ldots,\omega^n\frac{d^n}{d\omega^n} \ln\Psi_{\beta,\gamma}(\omega)\right)
\end{multline}
as the general relationship between the normalized wavelet derivatives and the derivatives of the logarithm of the wavelet. Here $B_n\left(c_1,c_2,\ldots  c_n\right)$ is the $n$th-order complete Bell polynomial defined implicitly by (\ref{expcumulants}). We then find
\begin{eqnarray}
\widetilde\Psi_{1;\beta,\gamma}(\omega_{\beta,\gamma})&=& 0\\
\widetilde\Psi_{2;\beta,\gamma}(\omega_{\beta,\gamma})&=& -\beta\gamma\label{Morsederivative2}
\\\widetilde\Psi_{ 3;\beta,\gamma}(\omega_{\beta,\gamma})&=& -\beta\gamma (\gamma- 3)
\\\widetilde\Psi_{4;\beta,\gamma}(\omega_{\beta,\gamma})&=&3(\beta\gamma) ^ 2-\beta\gamma\left[(\gamma- 3)^2+2\right]\label{Morsederivative4}
\end{eqnarray}
as first few values of the normalized wavelet derivatives at the peak frequency  $\omega_{\beta,\gamma}$.

\section{   Morse wavelet energy and peak frequencies}\label{gammaappendix}
For the generalized Morse wavelets, the wavelet ``energy frequency'' $\widetilde \omega_{\beta, \gamma}$  defined in (\ref{energyfrequency}) is given by a ratio of gamma functions (\ref{Morseenergyfrequency}).  Here we investigate why $\widetilde \omega_{\beta, \gamma}$ and $\omega_{\beta, \gamma}$ should become indistinguishable for $\gamma= 3$ and $P_{\beta, \gamma}>1$, as was observed in Figure~\ref{morsie_frequencies}(d).

Note that the asymptotic behavior of the gamma function is [\citen{abramowitz72},\;eqn.~6.1.39]
\begin{equation}
\Gamma\left(az+b\right)\sim  \sqrt{2\pi} e^ {-az} (az) ^ {az+b-1/2},\,\,\,|z|\longrightarrow\infty
\end{equation}
with $\left|\arg z\right|\leq\pi$ and $a>0$; here  $f(z)\sim g(z),\,|z|\longrightarrow\infty$ denotes $\lim_{|z|\longrightarrow\infty}f(z)/ g(z)= 1$ as usual. It follows that
\begin{eqnarray}
\frac{1}{x ^ {(n-1)r}}\frac{\Gamma\left(x+nr\right)}{\Gamma\left(x+r\right)}&\sim  &  1,\,\,\,|x|\longrightarrow\infty\label{gammaapproximation}
\end{eqnarray}
for real and positive $x$, $n$, and  $r$.  Choosing  $x= 2\beta/\gamma$ and $n= 2$, one obtains
\begin{equation}
\frac{\widetilde \omega_{\beta, \gamma}}{\omega_{\beta, \gamma}}=\frac{1}{\left(2\beta/\gamma\right)^ {1/\gamma}}\frac{\Gamma\left(2\beta/\gamma+2/\gamma\right)}{\Gamma\left(2\beta/\gamma+1/\gamma\right)}\sim1
,\,\,\, \beta\longrightarrow\infty
\label{Morsegammaapproximation}
\end{equation}
with fixed $\gamma$ but not, one may note, as $\gamma\longrightarrow 0$ with fixed $\beta$. Evaluating the left-hand side of (\ref{gammaapproximation}) for $n= 2$ numerically (not shown), one finds that for $r=1/3$, corresponding to the case $\gamma= 3$, this ratio in fact remains very close to unity for all $x\geq1$ and rapidly approaches its asymptotic value as $x$ increases. The minimum departure of the left-hand side of (\ref{gammaapproximation}) from unity at a particular value of $x$ is found to occur near $r=1/3$ for all $x\geq1$. Therefore the special properties of the $\gamma= 3$ wavelets have their origins in the behavior of the gamma function ratio in (\ref{gammaapproximation}) for $r=1/3$.
\section*{Acknowledgment}
We thank P.~J. Acklam for supplying Matlab code to implement the Dawson function.

\bibliography{ieeenames,spectral-003,ocean-008}

\begin{biography}{Jonathan M. Lilly (M'05)} was born in Lansing, Michigan, in 1972. He received the B.S. degree in Geology and Geophysics from Yale University, New Haven, Connecticut, in 1994, and the M.S. and Ph.D. degrees in Physical Oceanography from the University of Washington, Seattle, Washington, in 1997 and 2002, respectively.

He was a Postdoctoral Researcher at the University of Washington, Seattle, Applied Physics Laboratory and School of Oceanography from 2002 to 2003, and at the Laboratoire d'Oc$\acute{\mathrm{e}}$anographie Dynamique et de Climatologie of the Universit$\acute{\mathrm{e}}$ Pierre et Marie Curie, Paris, from 2003 to 2005. Since 2005 he has been a Research Associate at Earth and Space Research, a non-profit scientific institute in Seattle.  His research interests are oceanic vortex structures, satellite oceanography, time / frequency analysis methods, and wave-wave interactions.

Dr. Lilly is a member of the American Meteorological Society.

\end{biography}
\begin{biography}{Sofia C. Olhede (M'06)} was born in Spanga, Sweden in 1977. She received the M.Sci. and Ph.D. degrees in Mathematics from Imperial College London, London, U.K. in 2000 and 2003 respectively.

She held the posts of Lecturer (2002-2006) and Senior Lecturer (2006-2007)
at the Mathematics Department at Imperial College London, and in 2007 she
joined the Department of Statistical Science at University College London,
where she is professor of Statistics and honorary professor of Computer
Science. She serves on the Research Section of the Royal Statistical
Society, is a member of the Programme Committee of the International
Centre for Mathematical Sciences, and is an associate editor of the Journal
of the Royal Statistical Society, Series B (Statistical Methodology). Her
research interests include the analysis of diffusion weighted magnetic
resonance imaging data, complex-valued stochastic processes, and multiscale
methods.

Prof. Olhede is a fellow of the Royal Statistical Society and a member of the
Institute of Mathematical Statistics, the London Mathematical Society, and the Society of Industrial and Applied Mathematics.
\end{biography}

\end{document}